\documentclass[aps,pra,reprint,superscriptaddress]{revtex4-2}

\usepackage[english]{babel}
\usepackage{graphics}
\usepackage{array}
\usepackage{graphicx}
\usepackage{amsfonts,amssymb,amsmath}
\usepackage{color}
\usepackage{threeparttable}
\usepackage{longtable}
\usepackage[separate-uncertainty,per-mode=fraction]{siunitx}
\usepackage{booktabs}

\definecolor{darkblue}{rgb}{0.0,0.0,0.4}
\definecolor{darkgreen}{rgb}{0.0,0.4,0.0}
\definecolor{darkred}{rgb}{0.6,0.0,0.0}
\usepackage[bookmarksopen]{hyperref}
\hypersetup{colorlinks,linkcolor=darkblue,citecolor=darkblue,urlcolor=darkblue}

\newcolumntype{M}[1]{>{\centering\arraybackslash}m{#1}}
\newcommand{\Op}[1]{\boldsymbol{\hat{#1}}} 

\begin{document}

\title{Absorption spectroscopy for laser cooling and\\high-fidelity detection of barium monofluoride molecules}
\author{Marian Rockenh\"auser} 
\thanks{These two authors contributed equally}
\affiliation{5. Physikalisches  Institut  and  Center  for  Integrated  Quantum  Science  and  Technology,Universit\"at  Stuttgart,  Pfaffenwaldring  57,  70569  Stuttgart,  Germany}

\author{Felix Kogel}
\thanks{These two authors contributed equally}
\affiliation{5. Physikalisches  Institut  and  Center  for  Integrated  Quantum  Science  and  Technology,Universit\"at  Stuttgart,  Pfaffenwaldring  57,  70569  Stuttgart,  Germany}

\author{Einius Pultinevicius}
\affiliation{5. Physikalisches  Institut  and  Center  for  Integrated  Quantum  Science  and  Technology,Universit\"at  Stuttgart,  Pfaffenwaldring  57,  70569  Stuttgart,  Germany}

\author{Tim Langen}
\email{t.langen@physik.uni-stuttgart.de}

\affiliation{5. Physikalisches  Institut  and  Center  for  Integrated  Quantum  Science  and  Technology,Universit\"at  Stuttgart,  Pfaffenwaldring  57,  70569  Stuttgart,  Germany}

\begin{abstract} Currently, there is great interest in laser cooling of barium monofluoride (BaF) molecules for precision tests of fundamental symmetries. We use high-resolution absorption spectroscopy to characterize several as yet imprecisely known transition frequencies required to realize such cooling. We extract an improved set of molecular constants for the bosonic ${}^{138}\mathrm{Ba}^{19}$F and ${}^{136}\mathrm{Ba}^{19}$F isotopologues, confirm the existence of a significant hyperfine splitting in the excited state of the laser cooling cycle, and investigate the effects of this splitting on the achievable cooling forces. As a direct application of our spectroscopic insights, we experimentally demonstrate nearly background-free fluorescence imaging of a BaF molecular beam in a glass cell vacuum chamber. We expect such high-fidelity detection to be useful for various types of precision measurement scenarios.
\end{abstract}

\maketitle

\section{Introduction}
In recent years, laser cooling has revolutionized the field of molecular science~\cite{Fitch2021}. Starting from early proposals~\cite{DiRosa2004,Stuhl2008,Isaev2016}, several diatomic~\cite{Barry2014,Truppe2017,Anderegg2017,Collopy2018} and polyatomic~\cite{Baum2020,Vilas2021} molecular species can now routinely be cooled, trapped and studied at microkelvin temperatures. Currently, great efforts are being made to further expand the list of laser-coolable molecules, to include species that are specifically suited for certain applications ranging from precision measurements~\cite{Tarbutt2013,Kozyryev2017,Norrgard2017,Aggarwal2018,Altuntas2018,Lim2018,Rourke2019,GarciaRuiz2020} and ultracold chemistry~\cite{McNally2020,Mitra2020,Klos2020,Cheuk2020} to quantum simulation and information processing~\cite{Blackmore2018,Anderegg2019,Caldwell2020,Chae2021}. One such species is barium monofluoride (BaF)~\cite{Chen2017,Albrecht2020}, which has, due to its large mass and internal structure, promising applications in precision tests of fundamental symmetries~\cite{Aggarwal2018,Altuntas2018,Demille2008,Kogel2021,Vutha2018}. 

\begin{figure}[tb]
    \centering
    \includegraphics[width=0.42\textwidth]{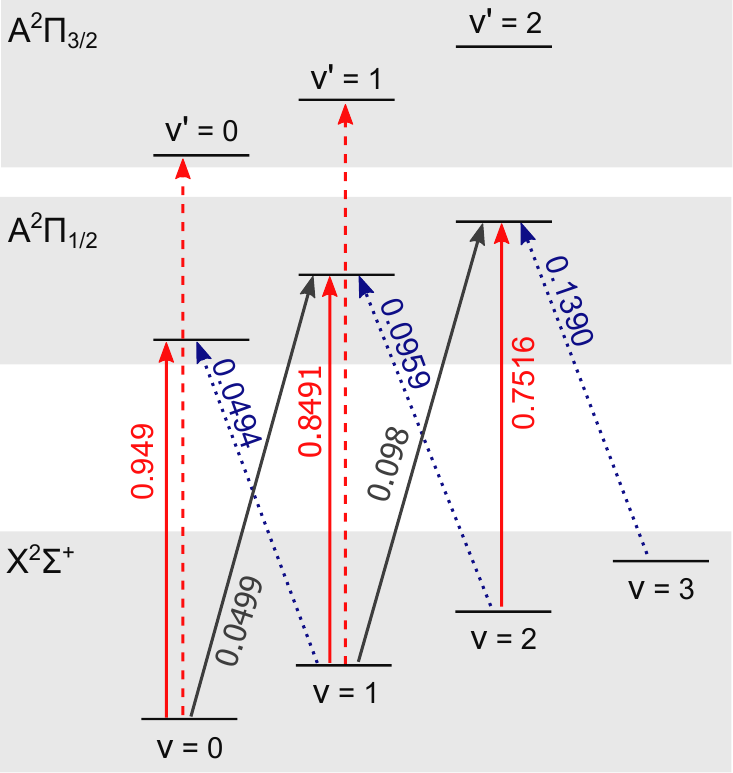}
    \caption{We study transitions between $\mathrm{X}^2\Sigma^+$ and $\mathrm{A}^2\Pi_{1/2}$  with $\Delta \nu=\nu'-\nu=0$ (red solid arrows, $\nu=0\rightarrow \nu'=0$ being the main \textit{cooling} transition), $\Delta \nu=-1$ (blue dotted arrows, \textit{repumping} transitions),  $\Delta \nu=+1$ (gray arrows, \textit{depumping} transitions), as well as transitions between $\mathrm{X}^2\Sigma^+$ and $\mathrm{A}^2\Pi_{3/2}$ with $\Delta \nu=0$ (red dashed arrows). The respective designations --- \textit{cooling}, \textit{repumping} and \textit{depumping} --- are motivated by laser cooling and imaging strategies based on these transitions, which are described further in the main text. The strength of each individual transition is determined by the population of the corresponding ground state, as set by the vibrational temperature (see App. A), and by their respective Franck-Condon factors (given next to the corresponding arrows, see App. B).}
    \label{fig:levelscheme}
\end{figure}

A simplified level scheme for the lowest-energy states of this species is shown in Fig.~\ref{fig:levelscheme}. A typical laser cooling strategy exploits the diagonal Franck-Condon factors between the $\mathrm{X}^2\Sigma^+$ and $\mathrm{A}^2\Pi_{1/2}$ electronic states, and combines the $\nu=0\rightarrow \nu'=0$  (main cooling) transition with the $\nu=1\rightarrow \nu'=0$, $\nu=2\rightarrow\nu'=1$, and \mbox{$\nu=3\rightarrow\nu'=2$} (1st, 2nd and 3rd repumping) transitions to realize a vibrationally quasi-closed optical cycle, where losses into higher vibrational states with quantum numbers $\nu>3$ are strongly suppressed.

However, it has recently been suggested that --- in contrast to similar laser-coolable monohalides such as CaF and SrF --- the $\mathrm{A}^2\Pi_{1/2}$ hyperfine structure exhibits a significant splitting, which could strongly affect the laser cooling forces~\cite{Denis2022,Marshall2022,Bu2022}. Moreover, while there exists an extensive body of spectroscopic data for BaF~\cite{Ryzlewicz1980,Ernst1986,Effantin1990,Bernard1990,Steimle2011,Bu2022}, the frequencies of the repumping transitions are so far not known precisely enough to realize laser cooling. While Doppler forces have already been demonstrated for BaF~\cite{Zhang2022}, applying them has thus resulted in most molecules being lost into unaddressed vibrational states, rather than being cooled. To successfully accumulate BaF molecules at low temperatures by laser cooling, further comprehensive high-resolution spectroscopy is thus required. 

\newpage
Motivated by this, we here study the lowest vibrational states of BaF molecules in a cryogenic buffer gas cell. In such a cell, hot molecules are created by laser ablation of a precursor, and subsequently cooled by collisions with a cold helium gas~\cite{Hutzler2012,Albrecht2020}. Buffer gas cooling is very efficient for rotational and translational degrees of freedom, but is known to be less efficient for vibrational degrees of freedom~\cite{Barry2011,Bu2017,Albrecht2020}. This leads to sizable populations in the lowest few vibrational states, which we probe via direct absorption spectroscopy. By investigating various rovibrational transitions between the $\mathrm{X}^2\Sigma^+$ electronic groundstate and the $\mathrm{A}^2\Pi_{1/2}$ and $\mathrm{A}^2\Pi_{3/2}$ excited states (see Fig.~\ref{fig:levelscheme}), we improve the relevant molecular constants for the most abundant bosonic isotopologues ${}^{138}\mathrm{Ba}^{19}$F and ${}^{136}\mathrm{Ba}^{19}$F. Our measurements confirm an important hyperfine splitting in the $\mathrm{A}^2\Pi_{1/2}$ state, allow us to determine the as yet unknown repumping wavelengths required for laser cooling, and facilitate nearly background-free fluorescence imaging of the molecular beam using other of the newly characterized transitions. 

\section{Experimental setup}
\begin{figure}[tb]
    \centering
    \includegraphics[width=0.48\textwidth]{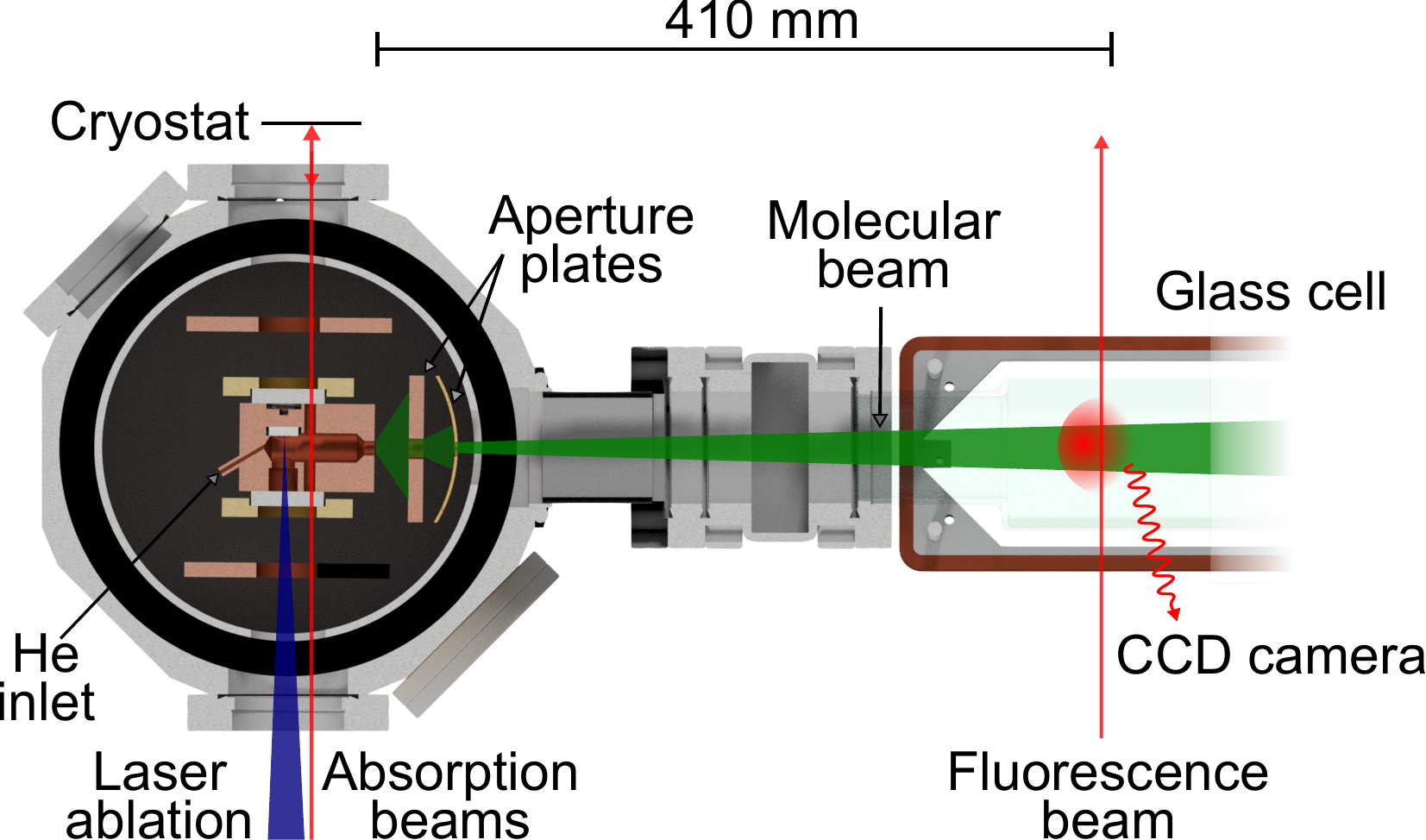}
    \caption{Experimental setup. BaF molecules are created via ablation of a solid $\mathrm{BaF}_2$ precursor inside a helium buffer gas cell, which is cooled by a 4\,K cryostat. We probe the molecules via absorption inside the cell using laser beams aligned parallel or anti-parallel with the ablation laser, or by a combination of these two beams to realize Doppler-free spectroscopy. The molecules exit the buffer gas cell via an aperture and form a slow molecular beam~\cite{Albrecht2020}, which can be probed further using fluorescence inside a glass cell vacuum chamber.}
    \label{fig:experiment}
\end{figure}

Our experiment starts with the ablation of solid $\mathrm{BaF}_2$ precurser target inside a buffer gas cell (see Fig.~\ref{fig:experiment}). The cryogenic cell is attached to the cold plate of a 4\,K helium cryostat. Helium gas with a flux of $0.6\,$sccm is introduced into the cell through a capillary. The helium atoms collide with the molecules formed via ablation and thermalize them to temperatures around 4\,K.  

We subsequently study the molecules inside the buffer gas cell using absorption spectroscopy. We use probe laser beams aligned either parallel or anti-parallel to the direction of the ablation laser to obtain coarse spectra, as well as a combination of two counter-propagating beams formed by retro-reflection to extract Doppler-free spectra using saturated absorption spectroscopy~\cite{Skoff2009}. The beams have diameters of $1\,$mm and are located approximately $1\,$cm downstream from the ablation spot. Their powers are actively stabilized by means of a fiberized amplitude modulator. These powers are approximately \SI{300}{\micro\watt} in the single beam configurations, and \SI{30}{\micro\watt} for the retro-reflected beam in the Doppler-free configuration.

To obtain spectra we scan the laser frequency and integrate over the time-resolved molecular absorption traces~\cite{Albrecht2020}. The time window for this integration is chosen such that translational and rotational degrees of freedom of the molecules have thermalized through collisions with the helium gas~\footnote{See Ref.~\cite{Wright2022} for a study of this effect in various other molecular species.}. To obtain the final absorption signal for a particular frequency, we average over several experimental runs and reduce statistical noise in the data with a moving average. 

We calibrate the absolute frequencies of the absorption laser for the scans using two diffferent wavemeters that we regularly validate against cesium and rubidium vapor spectroscopies, the known main cooling transition in ${}^{138}\mathrm{Ba}^{19}$F and a frequency stabilized helium neon laser. With this we conservatively estimate the absolute long-term accuracy of the observed frequencies to be better than $50\,$MHz, with the relative short-term accuracy being even much better.

A potential systematic uncertainty in our measurement could be caused by the residual motion of molecules in the direction opposite to the ablation laser following ablation. We exclude this possibility by comparing the absorption of individual beams co- and counter-propagating to the ablation direction, observing relative Doppler shifts between the two that are well below the above absolute accuracy of the frequency calibration. 

Consistent with these expected small uncertainties, we find excellent agreement with other recent measurements for the absolute transition wavelength of the $\mathrm{X}^2\Sigma^+(\nu=0)\rightarrow\mathrm{A}^2\Pi_{1/2}(\nu'=0)$ main cooling transition~\cite{Steimle2011,Chen2017,Albrecht2020}.

\begin{figure}[tb]
    \centering
    \includegraphics[width=0.47\textwidth]{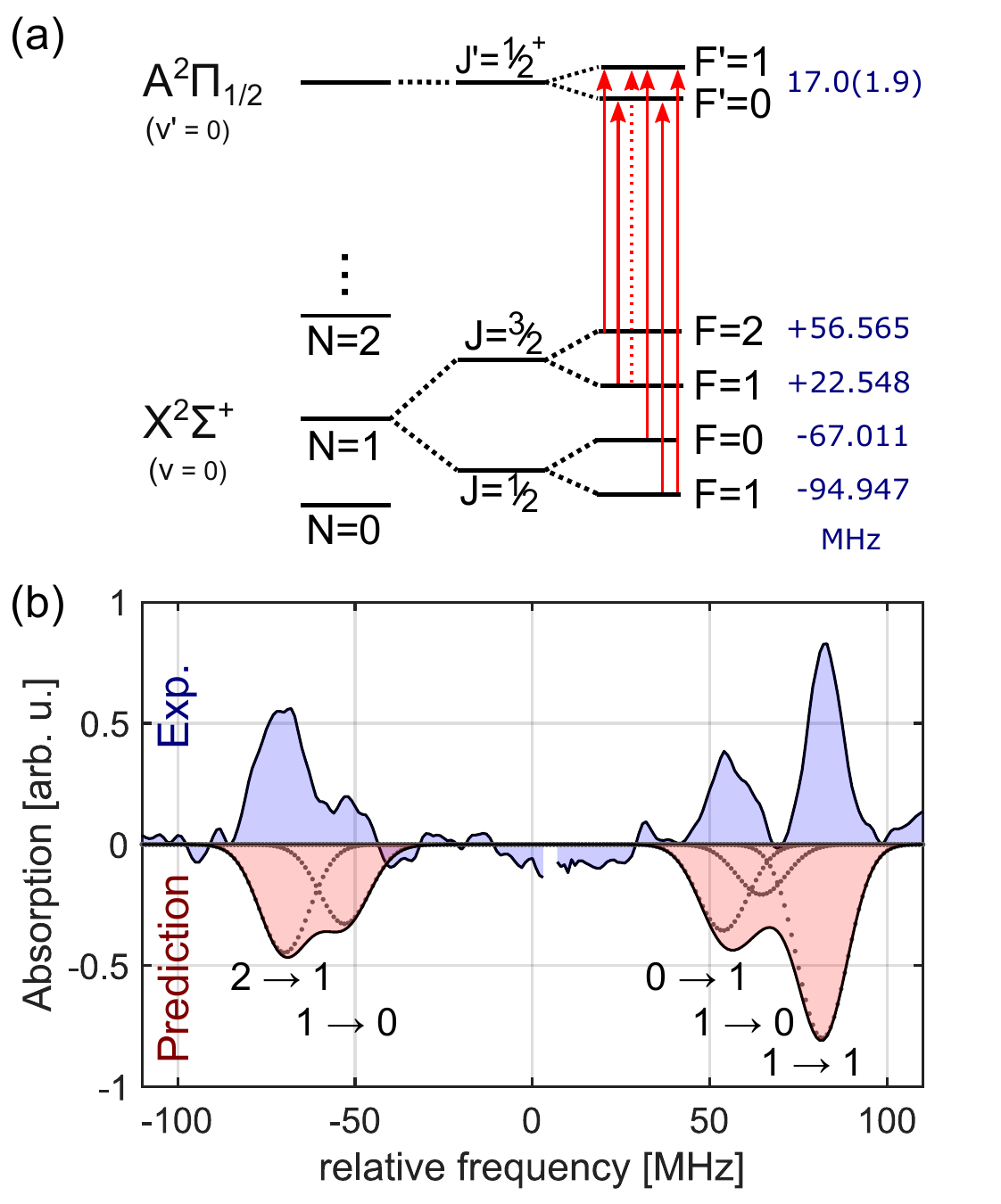}
    \caption{(a) Rotational and hyperfine substructure of the $\mathrm{X}^2\Sigma^+(\nu=0,N=1)\rightarrow\mathrm{A}^2\Pi_{1/2}(\nu'=0,J'=1/2^+)$ laser cooling transition in ${}^{138}\mathrm{Ba}^{19}$F~\cite{Mawhorter, Chen2016}. Following excitation from the $N=1$ state, parity selection rules only allow decay back from $J'=1/2^+$ to the initial state, realizing a rotationally closed cycle. In total six lines appear for this transition (arrows), one of which (dashed arrow) is nearly forbidden. (b) Saturated absorption spectroscopy of this transition. The experimental data has been averaged over $4$ absorption traces and is in excellent agreement with other recent measurements~\cite{Denis2022,Marshall2022,Bu2022}. The prediction is based on the molecular constants from Ref.~\cite{Denis2022} with a linewidth of $6\,$MHz. From fits of the peak positions, we extract an excited state splitting of $17.0(1.9)\,$MHz.} 
    \label{fig:dopplerfree}
\end{figure}

\begin{figure*}[tb]
    \centering
    \includegraphics[width=0.96\textwidth]{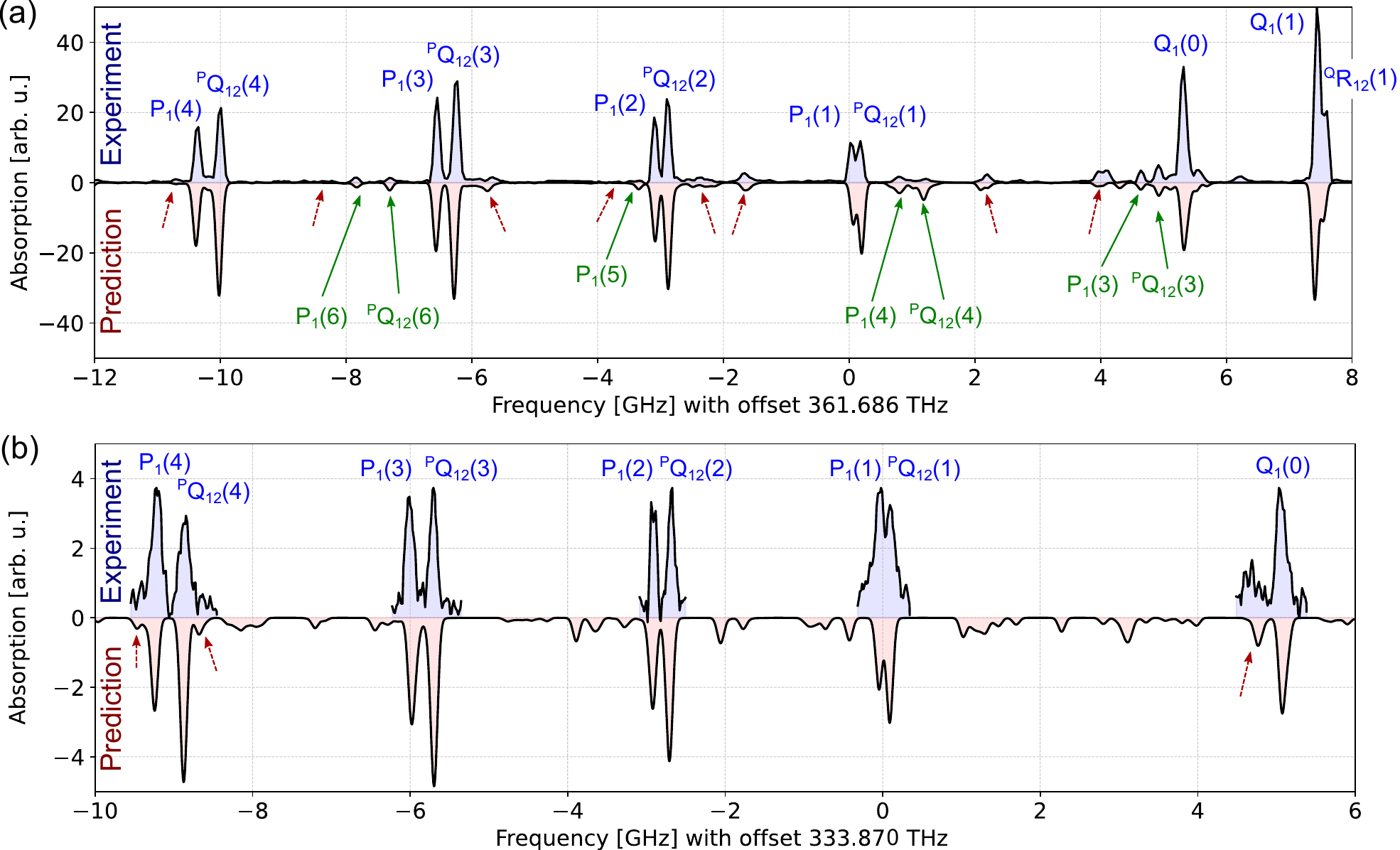}
   \caption{Example spectra of the (a) $\mathrm{X}^2\Sigma^+(\nu=0)\rightarrow\mathrm{A}^2\Pi_{1/2}(\nu'=1)$ and (b) $\mathrm{X}^2\Sigma^+(\nu=2)\rightarrow\mathrm{A}^2\Pi_{1/2}(\nu'=1)$  transitions. The top and bottom half show experimental data and predicted spectra including all isotopologues, respectively. Average absorption in (a) is typically around $20$ times lower than for the main cooling transition, due to the lower Franck-Condon factor. Absorption in (b) is further reduced by a factor of approximately $5$ due to lower population in $\nu=2$. No data in between the main transitions was recorded in (b). ${}^{138}\mathrm{Ba}^{19}$F transitions are labelled in blue and ${}^{136}\mathrm{Ba}^{19}$F transitions in green with arrows. Additional dashed red arrows indicate closely-spaced transitions of the fermionic ${}^{137}\mathrm{Ba}^{19}$F and ${}^{135}\mathrm{Ba}^{19}$F isotopologues, which will be analyzed in future work.\vspace{-8pt}}
    \label{fig:spectra}
\end{figure*}

\section{$\mathrm{A}^2\Pi_{1/2}$ hyperfine structure}
In a first measurement, we study the structure of the $J'=1/2^+$ sublevel of the $\mathrm{A}^2\Pi_{1/2}(\nu'=0)$ state in ${}^{138}\mathrm{Ba}^{19}$F. Here, $J'$ is the quantum number associated with the total angular momentum without nuclear spin, and the $+$ sign in the exponent denotes the state's positive parity. This state typically serves as the excited state for laser cooling, exploiting parity selection rules to rotationally close the optical cycle ~\cite{Albrecht2020,Fitch2021} (see Fig.~\ref{fig:dopplerfree}a). 

Recent work~\cite{Denis2022,Marshall2022,Bu2022} has suggested that the two hyperfine sublevels of this state, denoted by their total angular momentum quantum numbers $F'=0$ and $F'=1$, are split by a finite energy difference. Such a splitting, if not addressed by a suitable choice of laser frequencies, can result in a severe decrease of the force in laser cooling experiments. A precise characterization is thus essential for successful laser cooling of BaF. 

To this end, we record spectra of the $\mathrm{X}^2\Sigma^+(\nu=0,N=1)\rightarrow\mathrm{A}^2\Pi_{1/2}(\nu'=0,J'=1/2^+)$ transition in the single beam and retro-reflected beam configurations. While the former shows Doppler-broadened absorption lines with a full width at half maximum of around \SI{65}{\mega\hertz} --- far exceeding any expected hyperfine splitting --- the latter reveals Doppler-free features corresponding to the various hyperfine transitions in BaF. We account for broadening due to different powers of the two beams, normalize and substract the absorption signals, which yields the absorption lines shown in Fig.~\ref{fig:dopplerfree}b. These lines are in excellent agreement (root-mean-square deviation $1.64\,$MHz) with predictions based on recently improved excited state hyperfine constants~\cite{Denis2022,Marshall2022}. From their relative locations and the known ground-state hyperfine splittings~\cite{Mawhorter,Chen2016} we deduce an excited state hyperfine splitting of $17.0(1.9)\,$MHz between $F'=0$ and $F'=1$, with the uncertainty being dominated by frequency uncertainties between individual averaged experimental runs. This value agrees well with recent measurements finding a splitting of $17.7(2.1)\,$MHz~\cite{Bu2022} and $17.20(17)\,$MHz, respectively~\cite{Denis2022,Marshall2022}.

The observed splitting can be included in multi-level optical Bloch equations to model the laser cooling of BaF~\cite{Kogel2021}. We find that when using a conventional laser configuration with four equidistant laser sidebands~\cite{Chen2016,Albrecht2020}, the cooling forces are approximately halved due to the finite splitting (see App. C). Moreover, the splitting significantly reduces the velocity range in which cooling forces act. For efficient cooling, it is thus essential to take this splitting into account. These observations can be intuitively understood by the fact that the splitting causes the conventionally chosen laser sidebands to be blue detuned from some of the closely spaced transitions, while remaining red detuned from others. This leads to an interplay of heating and cooling forces, which greatly affects the efficiency of the overall cooling process. 

\section{Rovibrational spectroscopy}

In a second set of measurements, we use the single absorption beam configuration to extract spectra over a range of frequencies on the order of several GHz to observe various other rovibrational transitions (see Fig.~\ref{fig:levelscheme}). 

Two example results, for the $\mathrm{X}^2\Sigma^+(\nu=0)\rightarrow\mathrm{A}^2\Pi_{1/2}(\nu'=1)$ (first depumping transition) and the $\mathrm{X}^2\Sigma^+(\nu=2)\rightarrow\mathrm{A}^2\Pi_{1/2}(\nu'=1)$ (2nd repumping transition), respectively, are shown in Fig.~\ref{fig:spectra}. The spectra reveal multiple strong rotational transitions of the most abundant isotopologue ${}^{138}\mathrm{Ba}^{19}$F. In addition to this, also weaker lines from many other isotopologues are clearly visible. A full list of all observed lines in all recorded spectra is given in App. D. 

In the following, we describe the procedure for analyzing these spectra. 

In BaF, the lowest excited states $\mathrm{A}^2\Pi$, $\mathrm{B}^2\Sigma^+$ and $A'\Delta$ are well known to strongly perturb each other~\cite{Effantin1990}. However, given that only the lowest vibrational and rotational states are relevant for laser cooling we can neglect this perturbation and analyze our data by diagonalizing a standard effective Hamiltonian~\cite{Brown2003,Steimle2011,Chen2016} for each electronic ground $\mathrm{X}^2\Sigma^+(\nu)$ and excited state $\mathrm{A}^2\Pi_{1/2}(\nu')$.

All couplings between the different angular momenta described by this Hamiltonian can be decoupled from the molecular vibrations whose energy is expressed by an anharmonic oscillator. The energies of the molecular states can thus be found by combining the eigenvalues of the effective Hamiltonian with an additional electronic-vibrational offset $T_\nu = T_{e} + \omega_e(\nu+1/2) - \omega_e\chi_e(\nu+1/2)^2+\omega_e y_e(\nu+1/2)^3$ where $T_e$ is the energy of the electronic state manifold, and $\omega_e$, $\omega_e\chi_e$ and $\omega_e y_e$ are vibrational constants.

The $\mathrm{X}^2\Sigma^+$ ground state of BaF is described by Hund's case (b) where the spin-orbit coupling is much weaker than the rotational energy. The rotation $\Op{N}$ is first coupled with the electron spin $\Op{S}$ to form the total angular momentum without nuclear spin $\Op{J}$. Subsequently, the total angular momentum $\Op{F}$ is formed by coupling $\Op{J}$ to the nuclear spin $\Op{I}$. The rotational ladder scales with $B_\nu N(N+1) + D_\nu N^2(N+1)^2$, with the rotational constants $B_\nu = B_e - \alpha_e (\nu + 1/2)$ and $D_\nu \approx D_e$. Each rotational state with quantum number $N$ is split into two states $J=N+1/2$ and $J=N-1/2$ by spin-rotation coupling, with the energy splitting characterized by the spin-rotation constant $\gamma_\nu \approx\gamma_e$. These states split further into a total of four hyperfine states $F=N-1,N,N,N+1$. The energy splitting between these states depends on the hyperfine constants $b_F$ and $c$. Here, $J$ and $F$ denote the quantum number of $\Op{J}$ and  $\Op{F}$, respectively.  

The excited states $\mathrm{A}^2\Pi_{1/2}$ and $\mathrm{A}^2\Pi_{3/2}$ are described by Hund's case (a). The largest contribution to their energies arises from spin-orbit coupling, which separates the two states by an energy given by the spin-orbit constant $A_\nu=A_e+\alpha_A(\nu+1/2)$. Note that $A_e$ is isotopically independent on the level of our precision~\cite{Drouin2001}. For each of these states there is again a ladder of rotational states, now characterized by $J$, as $N$ is no longer a good quantum number in Hund's case (a). Furthermore, due to $\Lambda$-doubling, for every $J$ there are two parity states, with their splitting characterized by the constant $p+2q$. An example for one of these states is the $\mathrm{A}^2\Pi_{1/2}(\nu'=0,J'=1/2^+)$ state studied in detail in the previous section. The total angular momentum $F$ is formed by coupling rotation to the nuclear spin, resulting in two hyperfine states $F=J-1/2$ and $F=J+1/2$ for each value of $J$. Their splitting is characterized by the hyperfine constants $a$, $b$, $c$ and $d$.

It is custom to label the resulting rotational transitions using the notation ${}^{\Delta N}\Delta J_{a,b} (N)$, where $\Delta N = N' - N$ and
$\Delta J = J' - J$, with the primed and unprimed quantum
numbers referring to the excited and ground state,
respectively. The parameter $a$ is $1$ for transitions to the $\mathrm{A}^2\Pi_{1/2}$ and $2$ for transitions to the $\mathrm{A}^2\Pi_{3/2}$ states. The parameter $b$ is $1$ if the $J = N+1/2$ sublevel and $2$ if the $J = N-1/2$ sublevel of the given $\mathrm{X}^2\Sigma^+$ state is involved in the transition. In cases where $a = b$ or $\Delta J = \Delta N$ the repeated label is dropped. As usual, $O$, $P$, $Q$, $R$ and $S$ denote transitions where angular momentum quantum numbers change by $-2$, $-1$, $0$, $1$ and $2$, respectively. 

In total, we observe and assign $93$ lines. In particular, we locate the previously unknown second ($\nu=2\rightarrow\nu'=1$) and third ($\nu=3\rightarrow\nu'=2$) repumping transitions in ${}^{138}\mathrm{Ba}^{19}$F at $897.93162\,$nm and $900.17534\,$nm, respectively (see App. E). 

Our analysis starts with the strong lines associated with the most abundant ${}^{138}\mathrm{Ba}^{19}$F isotopologue. Once assigned, we extract line centers and fit them using the effective Hamiltonian. From the resulting constants for ${}^{138}\mathrm{Ba}^{19}$F we derive preliminary constants for ${}^{136}\mathrm{Ba}^{19}$F by mass scaling relations~\cite{Drouin2001, Doppelbauer2022}. This allows us to assign many of the weaker lines from this less abundant isotopologue. We then repeat the fitting procedure to obtain the final set of constants for ${}^{136}\mathrm{Ba}^{19}$F. 

The results of our analysis are summarized in Tab. \ref{table:constants}. A prediction based on these results is in good agreement with our experimental observations, as highlighted by the comparison in Fig.~\ref{fig:spectra}. We find the root-mean-square deviation of our predicted line centers from the experimental data to be less than $21\,$MHz. This agreement is an order of magnitude better than the agreement with predictions based on previously extracted constants~\cite{Effantin1990}. 

\begin{table}[tb]
\centering
\begin{threeparttable}
\caption{Molecular constants in cm$^{-1}$ for ${}^{138}\mathrm{Ba}^{19}$F and ${}^{136}\mathrm{Ba}^{19}$F. Additional constants for ${}^{137}\mathrm{Ba}^{19}$F and ${}^{135}\mathrm{Ba}^{19}$F used for the predictions shown in Figs.~\ref{fig:dopplerfree} and~\ref{fig:spectra} were taken from previous work~\cite{Steimle2011,Mawhorter} and, where unknown, approximated via mass scalings~\cite{Drouin2001,Doppelbauer2022}.}

\begin{tabular}{M{0.15\textwidth} M{0.15\textwidth} M{0.15\textwidth}}
     \toprule
     Parameter & ${}^{138}\mathrm{Ba}^{19}$F& ${}^{136}\mathrm{Ba}^{19}$F\\
     \midrule
     \multicolumn{1}{l}{$\mathrm{X}^2\Sigma^+$}\\ 
     $\omega_e$          & 469.4052(34)      & 469.8289(26)\\
     $\omega_e \chi_e$   &  1.8346(21)        & 1.8402(13)\\
     $10^3\times\omega_e y_e$ & 3.06(37)     & 3.07\tnote{d}\\
           
     $B_e$           & 0.216529655(19)\tnote{a} & 0.216915616(19)\tnote{a}\\
     $10^7\times D_e$    & 1.84299(23)\tnote{a}     & 1.84956(23)\tnote{a}\\
     $10^3\times\alpha_e$& 1.163568(42)\tnote{a} & 1.166680(42)\tnote{a}\\
     $10^3\times\gamma_e$& 2.69905(12)\tnote{a} & 2.70386(12)\tnote{a}\\
     $10^3\times b_F$    & 2.196510(16)\tnote{a} & 2.196510(16)\tnote{a}\\
     $10^3\times c$      & 0.243647(44)\tnote{a} & 0.243647(44)\tnote{a}\\
     \addlinespace 
     \multicolumn{1}{l}{$\mathrm{A}^2\Pi$}\\ 
     $T_e$               & 11962.0524(23)    & 11962.0586(18)\\
     $\omega_e$          & 437.9536(18)      & 438.3443(21)\\
     $\omega_e \chi_e$   & 1.8714(5)         & 1.8747(11)\\
     $A_e$               & 632.5369(25)      & 632.5369(25)\tnote{d}\\
     $\alpha_A$          & -0.5102(19)       & -0.5102(7)\\
     $A_{00}$            & 632.28175\tnote{b} & \\
     $B_e$               & 0.21227(9)          & 0.21274(12)\\
     $10^7\times D_e$    & 2.00\tnote{b}      & 2.01\tnote{d}\\
     $10^3\times\alpha_e$& 1.12(7)          & 1.23(10)\\
     $p+2q$              & -0.257550\tnote{b} & -0.2580091\tnote{d}\\
     $10^3\times a$      & 0.8856(33)\tnote{c} & 0.8856(33)\tnote{d}\\
     $10^3\times (b+c)$  & -0.1848(63)\tnote{c} & -0.1848(63)\tnote{d}\\
     $10^3\times d$      & 0.1194(47)\tnote{c} & 0.1194(47)\tnote{d}\\
     \bottomrule
\end{tabular}

\label{table:constants}
\begin{tablenotes}
\item[a] Reference~\cite{Mawhorter}.
\item[b] Reference~\cite{Steimle2011}.
\item[c] Reference~\cite{Denis2022}.
\item[d] Value obtained from scaling the value for ${}^{138}\mathrm{Ba}^{19}$F.
\end{tablenotes}
\end{threeparttable}
\end{table}

\begin{figure}[tb]
    \centering
    \includegraphics[width=0.41\textwidth]{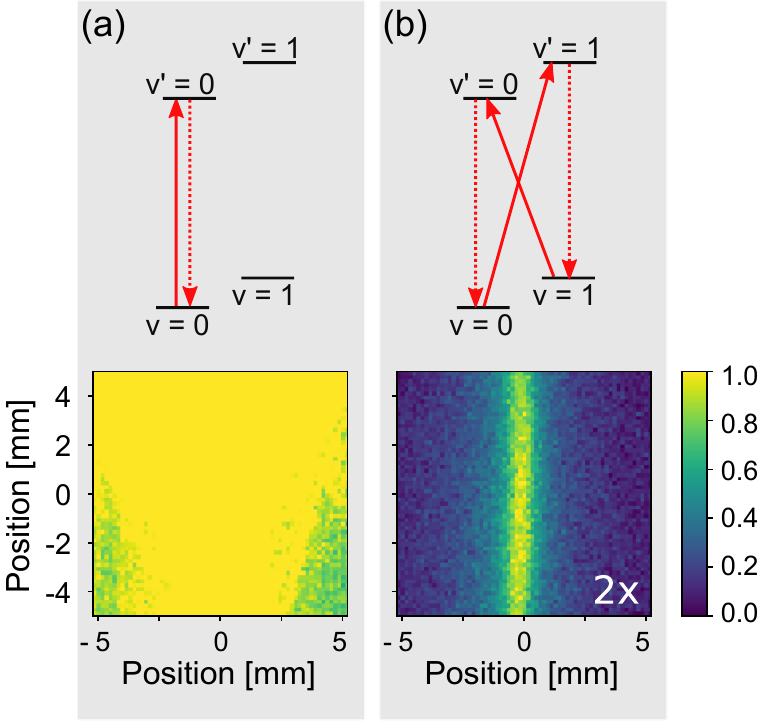}
    \vspace{10pt}
    \caption{Resonant Raman optical cycling imaging on the $\mathrm{X}^2\Sigma^+(\nu)$ to $\mathrm{A}^2\Pi_{1/2}(\nu')$ transition. (a) Conventional fluorescence imaging on the cooling transition ($\nu=0 \rightarrow \nu'=0$). Following excitation (solid arrow), the fluorescence is mainly emitted on the same transition (dashed arrow). This makes it challenging to record the molecular fluorescence in situations where stray light and reflections can not be suppressed. In this example, we illuminate a molecular beam in an uncoated glass cell, which leads to almost uniform saturation of the camera used to record the fluorescence. The color bar denotes normalized photon counts per pixel. (b) This challenge can be addressed by driving the first repump ($\nu=1 \rightarrow \nu'=0$) and depump ($\nu=0 \rightarrow \nu'=1$) transitions simultaneously (solid arrows), resulting in spectrally well-separated fluorescence on the cooling ($\nu=0 \rightarrow \nu'=0$) and $\nu=1 \rightarrow \nu'=1$ transitions (dashed arrows) and, hence,  a nearly background-free imaging of the molecular beam. Both images are averages over $40$ experimental realizations. For clarity, the image in (b) has been scaled by a factor of 2.}
    \label{fig:bowtie}
\end{figure}

\section{Fluorescence imaging}
As an application and a further check for our absorption spectroscopy, we demonstrate nearly background-free fluorescence imaging of the molecular beam, as it exits the buffer gas cell and passes through the uncoated part of a glass cell vacuum chamber (see Fig.~\ref{fig:experiment}). 

To do so, we employ resonant Raman optical cycling imaging~\cite{Shaw2021}. In this technique, a vibrationally quasi-closed optical cycle is realized by simultaneously driving both the first repumping ($\nu=1\rightarrow \nu'=0$) and depumping ($\nu=0\rightarrow \nu'=1$) transitions. While the scattering rate, and hence any potential cooling forces, obtained using this optical cycle are lower than in the conventional laser cooling cycle, this approach yields fluorescence on the first two $\Delta \nu=0$ transitions ($\nu=0\rightarrow \nu'=0$ and $\nu=1\rightarrow \nu'=1$, respectively). This fluorescence is spectrally well-separated from the laser light exciting the molecules by several tens of nanometers. Consequently, band pass filters can be used to effectively reduce stray light affecting the images. 

The results of the imaging procedure are shown in Fig.~\ref{fig:bowtie}. We use fluorescence beam diameters of $1\,$mm and the same power of $250\,$mW for the cooling beam, and the sum of repumping and depumping beam power, respectively. All beams are located $410\,$mm downstream from the exit aperture of the buffer gas cell. As for laser cooling, rotational branching is suppressed by parity selection rules and all beams include sidebands to address the various hyperfine components (see Fig.~\ref{fig:dopplerfree}). A magnetic field of $5$\,G is applied to remix Zeeman dark states. We find the transition frequencies in the beam to be identical with the ones determined above via in-cell absorption to within our measurement precision. 

As expected, driving the molecules on the cooling transition, as is conventionally done to realize optical cycling and laser cooling~\cite{Fitch2021}, leads to a large amount of stray light in the glass cell. This stray light saturates the camera used to record the fluorescence and makes it impossible to observe the molecular beam. When we instead drive the molecules using repumping and depumping light simultaneously, in combination with spectral filtering, a clear image of the molecular beam is observed. 

As the system formed by the combination of repumping and depumping light forms a closed optical cycle, the signal-to-noise ratio of this image could be further enhanced by extending the interaction time using a larger diameter laser beam. However, we note that in comparison to a previous implementation of this technique in SrF~\cite{Shaw2021}, the enhancement is limited in BaF by the slightly less favorable Franck-Conden factors of the $\mathrm{A}^2\Pi_{1/2}(\nu'=2)$ state, which result in stronger losses into higher vibrational states with $\nu\geq 3$. Nevertheless, even in this case, a striking enhancement of the signal-to-noise ration is obtained, which is readily transferable to less abundant BaF isotopologues and many other molecular species. The technique is thus an example for how the complex molecular structure, while making laser cooling challenging, can also be beneficial in ways that are not easily possible in simpler systems such as atoms. We expect such nearly background-free imaging to be particularly useful for detection protocols in precision experiments testing fundamental symmetries~\cite{Aggarwal2018,Altuntas2018}.

\section{Conclusion}
We have performed absorption spectroscopy of BaF molecules in a buffer gas cell. This has allowed us to identify the missing vibrational repumping transitions for laser cooling, as well as to resolve an important excited state hyperfine splitting in the $\mathrm{A}^2\Pi_{1/2}$ state, which has a strong influence on the obtainable laser cooling forces. 

By using a more precise frequency comb or wavemeter, the precision of our study could easily be pushed further, to a level where it becomes relevant for searches for the variation of fundamental constants~\cite{Kajita2018}, the realization of molecular clocks and THz metrology~\cite{Leung2022,Barontini2021}, or King-plot analysis of nuclear and molecular structure~\cite{Athanasakis2023}. Moreover, we also observe a large number of transitions of the odd, fermionic isotopologues ${}^{135}\mathrm{Ba}^{19}$F and ${}^{137}\mathrm{Ba}^{19}$F, which are relevant for parity violation studies~\cite{Altuntas2018,Kogel2021}. These transitions will be analyzed in future work. 

\section*{Acknowledgments}
We are indebted to Tilman Pfau for generous support. We thank Ralf Albrecht for contributions in the early stage of the experiment, and Timothy Steimle, Richard Mawhorter and Ed Grant for fruitful discussions. This project has received funding from the European Research Council (ERC) under the European Union’s Horizon 2020 research and innovation programme (Grant agreements No. 949431), Vector Stiftung, the RiSC programme of the Ministry of Science, Research and Arts Baden-W\"urttemberg and Carl Zeiss Foundation.
\section*{Appendix}
\subsection{Vibrational temperature}
\begin{figure}[htb]
    \centering
    \includegraphics[width=0.42\textwidth]{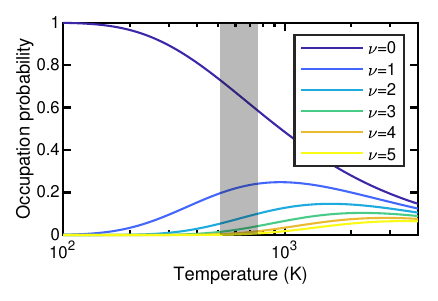}
    \caption{Occupation probability of the lowest vibrational levels of BaF as a function of temperature. The shaded bar indicates the estimated vibrational temperature of around $620\pm100\,$K, as evaluated from the observed peak absorption of the various transitions.}
    \label{fig:appendix_vibrations}
\end{figure}

In Fig.~\ref{fig:appendix_vibrations} we show the occupation probability of the vibrational states of BaF as a function of temperature. When combined with the Franck-Condon factors, this determines the strength of the vibrational transitions. In our experiments, the observed peak absorption values point towards a vibrational temperature in the range of several hundred Kelvin, which significantly exceeds the temperature of the buffer gas. In contrast to this, the translational and rotational degrees of freedom are well thermalized to temperatures of only a few Kelvin~\cite{Albrecht2020}. 

\subsection{Franck-Condon factors}
Based on the improved set of molecular constants obtained in this work, we follow the procedure outline in Ref.~\cite{Albrecht2020} to re-evaluate the theoretically expected Franck-Condon factors of BaF. The results are highlighted in Fig.~\ref{fig:levelscheme} and summarized in the tables below.\\~\\

    $A^2\Pi_{1/2}(\nu')\rightarrow X^2\Sigma(\nu'')$ (${}^{138}\textrm{Ba}^{19}$F)
    {\centering
    \begin{tabular*}{0.45\textwidth}{c @{\extracolsep{\fill}} cccc}
        \toprule
         & $\nu'=0$ & $\nu'=1$ & $\nu'=2$ & $\nu'=3$ \\ 
        \midrule
        $\nu''=0$ & $0.94895$ & $0.04994$ & $1 \times 10^{-3}$ & $6 \times 10^{-6}$\\ 
        $\nu''=1$ & $0.04939$ & $0.84910$ & $0.09795$ & $3 \times 10^{-3}$\\ 
        $\nu''=2$ & $1\times 10^{-3}$ & $0.09588$ & $0.75160$ & $0.14333$ \\ 
        $\nu''=3$ & $1\times10^{-5}$ & $5\times 10^{-3}$ & $0.13901$ & $0.65733$ \\ 
        \bottomrule
        \end{tabular*} 

        }

\vspace{19pt}
~\\
    $A^2\Pi_{1/2}(\nu')\rightarrow X^2\Sigma(\nu'')$ (${}^{136}\textrm{Ba}^{19}$F)
    {\centering
    \begin{tabular*}{0.45\textwidth}{c @{\extracolsep{\fill}} cccc}
        \toprule
         & $\nu'=0$ & $\nu'=1$ & $\nu'=2$ & $\nu'=3$ \\ 
        \midrule
        $\nu''=0$ & $0.95545$ & $0.05569$ & $1 \times 10^{-3}$ & $8 \times 10^{-6}$\\ 
        $\nu''=1$ & $0.04327$ & $0.85631$ & $0.10927$ & $4 \times 10^{-3}$\\ 
        $\nu''=2$ & $1\times 10^{-3}$ & $0.08411$ & $0.75936$ & $0.15997$ \\ 
        $\nu''=2$ & $1\times10^{-5}$ & $4\times 10^{-3}$ & $0.12210$ & $0.66549$ \\ 
        \bottomrule
    \end{tabular*}
\vspace{5pt}~\\~\\
    }

\subsection{Influence of the hyperfine splitting\\on the laser cooling forces}

To study the effect of the observed splitting between the $\mathrm{A}^2\Pi_{1/2} (\nu'=0,J'=1/2^+,F=0)$ and $\mathrm{A}^2\Pi_{1/2} (\nu'=0,J'=1/2^+,F=1)$ hyperfine states, we use multi-level Bloch equations to simulate the resulting laser cooling forces~\cite{Kogel2021,Kogel2021code}.

\begin{figure}[bht]
    \centering
    \hspace{-9pt}\includegraphics[width=0.49\textwidth]{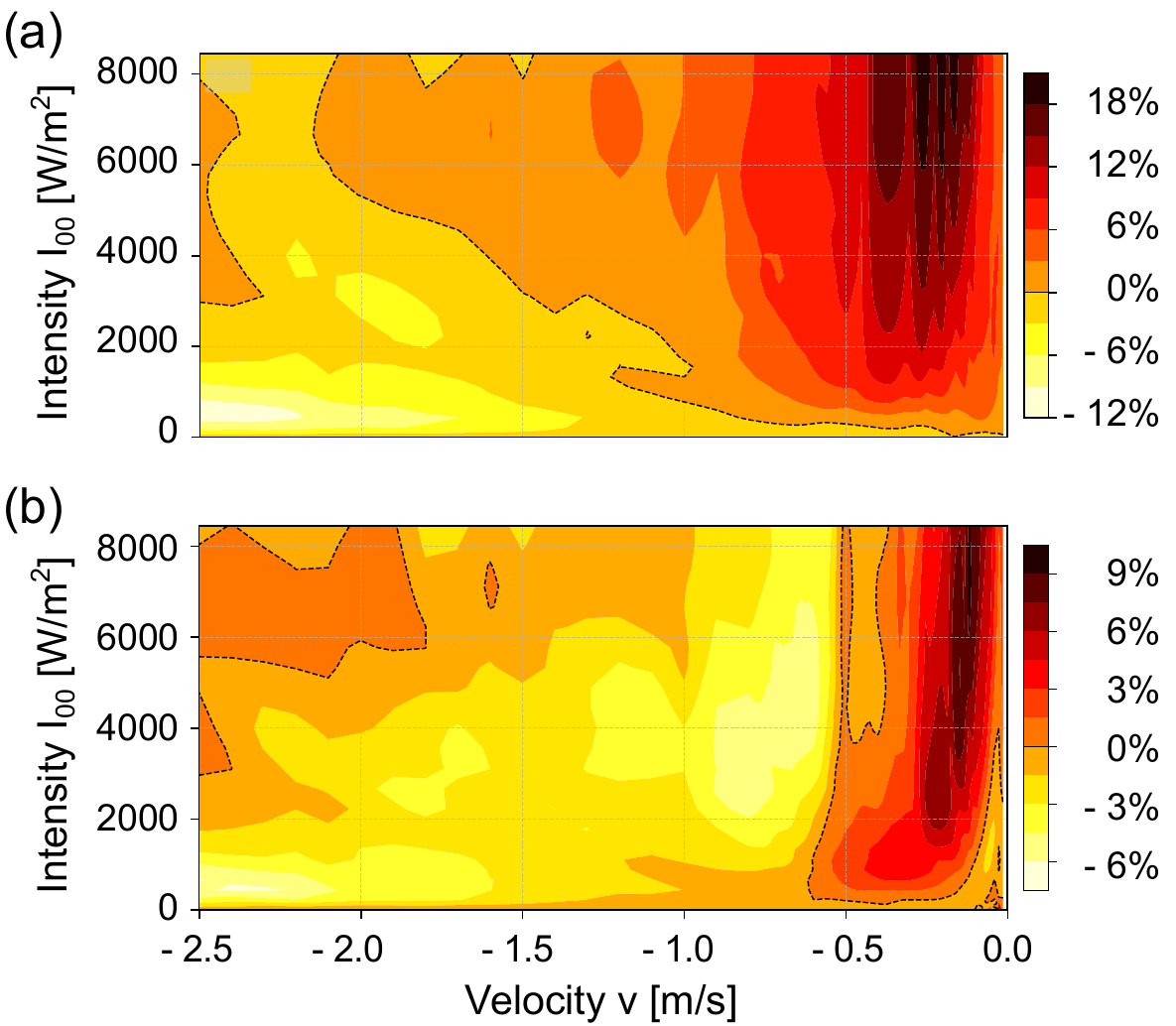}
    \caption{Laser cooling force profiles for a configuration with equidistant sidebands (a) with no excited state hyperfine splitting and (b) including hyperfine splitting. The sideband configuration is equivalent to the one generated by a single EOM operated at $39.33\,$MHz. Including the hyperfine splitting leads to a notable decrease in both magnitude and range of the laser cooling force (indicated by dashed lines), as the sidebands are no longer properly detuned from the individual transitions. The colorbar denotes the magnitude relative to the ideal value of $\hbar k\Gamma/2$ of a two-level system with the same wavelength. The intensity $I_{00}$ corresponds to the total intensity of all sidebands of a single laser beam that is retroreflected.}
    \label{fig:appendix_forces}
\end{figure}

As shown in Fig.~\ref{fig:dopplerfree}a, for laser cooling of BaF, the relevant ground state features four almost equidistant hyperfine sublevels. A conventional scheme to address these sublevels~\cite{Chen2016,Bu2017,Albrecht2020} thus relies on a single electro-optical modulator to imprint the required four laser sidebands. For zero exited state hyperfine splitting these sidebands almost perfectly match the four possible transitions. 

In contrast to this, for a finite excited state splitting there are in total six possible transitions. However, one of these transitions is nearly forbidden. Another one is comparably weak and overlaps with two stronger ones (see Fig.~\ref{fig:dopplerfree}a). A total number of four sidebands is thus still sufficient to address all ground state levels.

The resulting forces for both situations are compared in Fig.~\ref{fig:appendix_forces} as a function of employed laser intensity and the velocity of the molecules. In these simulations, molecules are located in the center of a pair of linearly polarized laser beams. The powers in the individual sidebands are $22\,\%$, $27\,\%$, $27\,\%$, and $22\,\%$ of the total power, respectively. We use a blue detuning of $1\,\Gamma$ and a comparably low magnetic field of $1\,$G, tilted by $60^{\circ}$ with respect to the laser beam polarization axis of the laser, such that magnetically-assisted Sisyphus cooling is the main cooling mechanism~\cite{Kogel2021}. As discussed in the main text, both the absolute magnitude and velocity range of the laser cooling force are significantly reduced if the excited state splitting is included. 

Given the comparably low scattering rate (excited state linewidth $\Gamma = 2\pi\times 2.83\,$MHz), high mass and low recoil velocity, properly chosen, arbitrarily spaced sideband frequencies will thus be essential to realize laser cooling of BaF. Such sidebands are typically generated using a combination of electro-optic and acousto-optic modulators. Finding the best number and location for these sidebands remains a complex optimization problem, which we will address in future work.

\subsection{List of observed transitions}
We assign a total number of $93$ lines, $64$ for ${}^{138}\mathrm{Ba}^{19}$F and $29$ for ${}^{136}\mathrm{Ba}^{19}$F. We denote the difference between the experimentally observed transition frequency and the prediction based on the constants in Tab.~\ref{table:constants} by $\Delta$. The uncertainty of the frequency for all observed transitions is conservatively estimated to be $50\,$MHz.

{\centering
\begin{longtable}{c c  c  c  r  c c}
    \toprule
     Ba isotope & Ex. state & $\nu$ & $\nu'$ & Transition & Freq. (THz) & $\Delta$ (MHz)\\
     \midrule
    \endfirsthead 
    138 & $^2\Pi_{1/2}$ & 0 & 0 & $Q_1$(0) & 348.666366 & -25\\
    138 & $^2\Pi_{1/2}$ & 0 & 0 & $^PQ_{12}$(1) & 348.661273 & 19\\
    138 & $^2\Pi_{1/2}$ & 0 & 0 & $P_1$(1) & 348.661140 & 21\\
    138 & $^2\Pi_{1/2}$ & 0 & 0 & $^PQ_{12}$(2) & 348.658303 & 14\\
    138 & $^2\Pi_{1/2}$ & 0 & 0 & $P_1$(2) & 348.658096 & 15\\
    138 & $^2\Pi_{1/2}$ & 0 & 0 & $^PQ_{12}$(3) & 348.655076 & 6\\
    138 & $^2\Pi_{1/2}$ & 0 & 0 & $P_1$(3) & 348.654760 & -23\\
    138 & $^2\Pi_{1/2}$ & 0 & 0 & $^PQ_{12}$(4) & 348.651572 & -22\\
    138 & $^2\Pi_{1/2}$ & 0 & 0 & $P_1$(4) & 348.651214 & -13\\
    \midrule
    138 & $^2\Pi_{1/2}$ & 1 & 0 & $Q_1$(0) & 334.703685 & 4\\
    138 & $^2\Pi_{1/2}$ & 1 & 0 & $^PQ_{12}$(1) & 334.698597 & -16\\
    138 & $^2\Pi_{1/2}$ & 1 & 0 & $P_1$(1) & 334.698461 & -18\\
    138 & $^2\Pi_{1/2}$ & 1 & 0 & $^PQ_{12}$(2) & 334.695779 & -9\\
    138 & $^2\Pi_{1/2}$ & 1 & 0 & $P_1$(2) & 334.695567 & -13\\
    138 & $^2\Pi_{1/2}$ & 1 & 0 & $^PQ_{12}$(3) & 334.692791 & 12\\
    138 & $^2\Pi_{1/2}$ & 1 & 0 & $P_1$(3) & 334.692507 & 16\\
    138 & $^2\Pi_{1/2}$ & 1 & 0 & $^PQ_{12}$(4) & 334.689593 & 11\\
    138 & $^2\Pi_{1/2}$ & 1 & 0 & $P_1$(4) & 334.689237 & 23\\
    \midrule
    138 & $^2\Pi_{1/2}$ & 1 & 1 & $Q_1$(0) & 347.728620 & 4\\
    138 & $^2\Pi_{1/2}$ & 1 & 1 & $^PQ_{12}$(1) & 347.723575 & 26\\
    138 & $^2\Pi_{1/2}$ & 1 & 1 & $P_1$(1) & 347.723440 & 26\\
    138 & $^2\Pi_{1/2}$ & 1 & 1 & $^PQ_{12}$(2) & 347.720612 & -5\\
    138 & $^2\Pi_{1/2}$ & 1 & 1 & $P_1$(2) & 347.720404 & -5\\
    138 & $^2\Pi_{1/2}$ & 1 & 1 & $^PQ_{12}$(3) & 347.717399 & -31\\
    138 & $^2\Pi_{1/2}$ & 1 & 1 & $P_1$(3) & 347.717111 & -31\\
    138 & $^2\Pi_{3/2}$ & 1 & 1 & $Q_{21}$(1) & 366.665911 & 3\\
    138 & $^2\Pi_{3/2}$ & 1 & 1 & $R_2$(1) & 366.666041 & 7\\
    138 & $^2\Pi_{3/2}$ & 1 & 1 & $Q_2$(2) & 366.640329 & 12\\
    138 & $^2\Pi_{3/2}$ & 1 & 1 & $P_{21}$(2) & 366.640092 & -21\\
    138 & $^2\Pi_{3/2}$ & 1 & 1 & $Q_2$(3) & 366.633292 & -3\\
    138 & $^2\Pi_{3/2}$ & 1 & 1 & $P_{21}$(3) & 366.633011 & 2\\
    \midrule
    138 & $^2\Pi_{1/2}$ & 0 & 1 & $Q_1$(0) & 361.691311 & -16\\
    138 & $^2\Pi_{1/2}$ & 0 & 1 & $^PQ_{12}$(1) & 361.686174 & -15\\
    138 & $^2\Pi_{1/2}$ & 0 & 1 & $P_1$(1) & 361.686032 & -23\\
    138 & $^2\Pi_{1/2}$ & 0 & 1 & $^PQ_{12}$(2) & 361.683115 & -3\\
    138 & $^2\Pi_{1/2}$ & 0 & 1 & $P_1$(2) & 361.682908 & -2\\
    138 & $^2\Pi_{1/2}$ & 0 & 1 & $^PQ_{12}$(3) & 361.679756 & 35\\
    138 & $^2\Pi_{1/2}$ & 0 & 1 & $P_1$(3) & 361.679448 & 14\\
    138 & $^2\Pi_{1/2}$ & 0 & 1 & $^PQ_{12}$(4) & 361.676002 & 6\\
    138 & $^2\Pi_{1/2}$ & 0 & 1 & $P_1$(4) & 361.675642 & 13\\
    \midrule
    138 & $^2\Pi_{1/2}$ & 2 & 1 & $Q_1$(0) & 333.875061 & -18\\
    138 & $^2\Pi_{1/2}$ & 2 & 1 & $^PQ_{12}$(1) & 333.870095 & 14\\
    138 & $^2\Pi_{1/2}$ & 2 & 1 & $P_1$(1) & 333.869968 & 21\\
    138 & $^2\Pi_{1/2}$ & 2 & 1 & $^PQ_{12}$(2) & 333.867318 & 29\\
    138 & $^2\Pi_{1/2}$ & 2 & 1 & $P_1$(2) & 333.867091 & 10\\
    138 & $^2\Pi_{1/2}$ & 2 & 1 & $^PQ_{12}$(3) & 333.864294 & -17\\
    138 & $^2\Pi_{1/2}$ & 2 & 1 & $P_1$(3) & 333.863995 & -29\\
    138 & $^2\Pi_{1/2}$ & 2 & 1 & $^PQ_{12}$(4) & 333.861139 & -5\\
    138 & $^2\Pi_{1/2}$ & 2 & 1 & $P_1$(4) & 333.860779 & 2\\
    \midrule
    138 & $^2\Pi_{1/2}$ & 1 & 2 & $^PQ_{12}$(1) & 360.636269 & 1\\
    138 & $^2\Pi_{1/2}$ & 1 & 2 & $P_1$(1) & 360.636124 & -9\\
    138 & $^2\Pi_{1/2}$ & 1 & 2 & $^PQ_{12}$(2) & 360.633243 & 14\\
    138 & $^2\Pi_{1/2}$ & 1 & 2 & $P_1$(2) & 360.633016 & -5\\
    138 & $^2\Pi_{1/2}$ & 1 & 2 & $^PQ_{12}$(3) & 360.629873 & 9\\
    138 & $^2\Pi_{1/2}$ & 1 & 2 & $P_1$(3) & 360.629575 & -2\\
    \midrule
    138 & $^2\Pi_{1/2}$ & 3 & 2 & $^PQ_{12}$(1) & 333.037912 & -40\\
    138 & $^2\Pi_{1/2}$ & 3 & 2 & $P_1$(1) & 333.037781 & -37\vspace{-3pt}\\
    138 & $^2\Pi_{1/2}$ & 3 & 2 & $^PQ_{12}$(2) & 333.035195 & -7\\
    138 & $^2\Pi_{1/2}$ & 3 & 2 & $P_1$(2) & 333.034983 & -11\\
    138 & $^2\Pi_{1/2}$ & 3 & 2 & $^PQ_{12}$(3) & 333.032326 & +55\\
    138 & $^2\Pi_{1/2}$ & 3 & 2 & $P_1$(3) & 333.032022 & +39\\
    \midrule
    138 & $^2\Pi_{1/2}$ & 2 & 2 & $Q_1$(0) & 346.787759 & -39\\
    138 & $^2\Pi_{1/2}$ & 2 & 2 & $^PQ_{12}$(1) & 346.782823 & 23\\
    138 & $^2\Pi_{1/2}$ & 2 & 2 & $P_1$(1) & 346.782675 & 9\\
    \midrule
    136 & $^2\Pi_{1/2}$ & 0 & 0 & $Q_1$(0) & 348.666087 & -23\\
    136 & $^2\Pi_{1/2}$ & 0 & 0 & $^PQ_{12}$(1) & 348.660993 & +29\\
    136 & $^2\Pi_{1/2}$ & 0 & 0 & $P_1$(1) & 348.660843 & +14\\
    136 & $^2\Pi_{1/2}$ & 0 & 0 & $P_1$(2) & 348.657792 & +5\\
    136 & $^2\Pi_{1/2}$ & 0 & 0 & $P_1$(3) & 348.654460 & -25\\
    136 & $^2\Pi_{1/2}$ & 0 & 0 & $P_1$(4) & 348.650892 & -33\\
    \midrule
    136 & $^2\Pi_{1/2}$ & 1 & 0 & $Q_1$(3) & 334.697248 & +0\\
    136 & $^2\Pi_{1/2}$ & 1 & 0 & $^QR_{12}$(3) & 334.697547 & +11\\
    136 & $^2\Pi_{1/2}$ & 1 & 0 & $Q_1$(4) & 334.698970 & +23\\
    136 & $^2\Pi_{1/2}$ & 1 & 0 & $^QR_{12}$(4) & 334.699312 & -3\\
    \midrule
    136 & $^2\Pi_{1/2}$ & 1 & 1 & $Q_1$(0) & 347.727482 & +12\\
    136 & $^2\Pi_{1/2}$ & 1 & 1 & $^PQ_{12}$(1) & 347.722378 & -15\\
    136 & $^2\Pi_{1/2}$ & 1 & 1 & $P_1$(1) & 347.722231 & -27\\
    136 & $^2\Pi_{1/2}$ & 1 & 1 & $^PQ_{12}$(2) & 347.719452 & -2\\
    136 & $^2\Pi_{1/2}$ & 1 & 1 & $P_1$(2) & 347.719244 & -1\\
    136 & $^2\Pi_{3/2}$ & 1 & 1 & $Q_{21}$(1) & 366.664762 & +2\\
    136 & $^2\Pi_{3/2}$ & 1 & 1 & $R_2$(1) & 366.664883 & -3\\
    136 & $^2\Pi_{3/2}$ & 1 & 1 & $Q_2$(2) & 366.639124 & +1\\
    136 & $^2\Pi_{3/2}$ & 1 & 1 & $P_{21}$(2) & 366.638919 & +0\\
    \midrule
    136 & $^2\Pi_{1/2}$ & 0 & 1 & $^PQ_{12}$(3) & 361.690927 & +9\\
    136 & $^2\Pi_{1/2}$ & 0 & 1 & $P_1$(3) & 361.690640 & +9\\
    136 & $^2\Pi_{1/2}$ & 0 & 1 & $^PQ_{12}$(4) & 361.687219 & +39\\
    136 & $^2\Pi_{1/2}$ & 0 & 1 & $P_1$(4) & 361.686808 & -5\\
    136 & $^2\Pi_{1/2}$ & 0 & 1 & $P_1$(5) & 361.682660 & -3\\
    136 & $^2\Pi_{1/2}$ & 0 & 1 & $^PQ_{12}$(6) & 361.678708 & -3\\
    136 & $^2\Pi_{1/2}$ & 0 & 1 & $P_1$(6) & 361.678169 & -13\\
    \midrule
    136 & $^2\Pi_{1/2}$ & 2 & 2 & $Q_1$(1) & 346.788038 & -0\\
    \midrule
    136 & $^2\Pi_{1/2}$ & 1 & 2 & $P_1$(4) & 360.635916 & +1\\
    136 & $^2\Pi_{1/2}$ & 1 & 2 & $^PQ_{12}$(5) & 360.632231 & -1\\
    \bottomrule
\end{longtable}}
\vspace{-2pt}
\subsection{Important transition wavelengths}
Below we summarize the wavelengths of the most important transitions for ${}^{138}\textrm{Ba}^{19}$F laser cooling and imaging. For all transitions  wavelengths quoted correspond to the center between the $^PQ_{12}(1)$ and $P_1(1)$ transitions.

{\centering
\begin{longtable}{c c  c  c l}
 \toprule
     Transition & $\nu$ & $\nu'$ & $\lambda$ (nm)  \\
     \midrule
    \endhead
 cooling & 0 & 0 &  859.83887\\
 1st repumper & 1 & 0 &  895.70892\\
 2nd repumper & 2 & 1 &  897.93162\\
 3rd repumper & 3 & 2 &  900.17534\\
 1st depumper & 0 & 1 &   828.87469\\
    \bottomrule
\end{longtable}

}

\bibliography{biblio}

\begin{thebibliography}{57}%
\makeatletter
\providecommand \@ifxundefined [1]{%
 \@ifx{#1\undefined}
}%
\providecommand \@ifnum [1]{%
 \ifnum #1\expandafter \@firstoftwo
 \else \expandafter \@secondoftwo
 \fi
}%
\providecommand \@ifx [1]{%
 \ifx #1\expandafter \@firstoftwo
 \else \expandafter \@secondoftwo
 \fi
}%
\providecommand \natexlab [1]{#1}%
\providecommand \enquote  [1]{``#1''}%
\providecommand \bibnamefont  [1]{#1}%
\providecommand \bibfnamefont [1]{#1}%
\providecommand \citenamefont [1]{#1}%
\providecommand \href@noop [0]{\@secondoftwo}%
\providecommand \href [0]{\begingroup \@sanitize@url \@href}%
\providecommand \@href[1]{\@@startlink{#1}\@@href}%
\providecommand \@@href[1]{\endgroup#1\@@endlink}%
\providecommand \@sanitize@url [0]{\catcode `\\12\catcode `\$12\catcode
  `\&12\catcode `\#12\catcode `\^12\catcode `\_12\catcode `\%12\relax}%
\providecommand \@@startlink[1]{}%
\providecommand \@@endlink[0]{}%
\providecommand \url  [0]{\begingroup\@sanitize@url \@url }%
\providecommand \@url [1]{\endgroup\@href {#1}{\urlprefix }}%
\providecommand \urlprefix  [0]{URL }%
\providecommand \Eprint [0]{\href }%
\providecommand \doibase [0]{https://doi.org/}%
\providecommand \selectlanguage [0]{\@gobble}%
\providecommand \bibinfo  [0]{\@secondoftwo}%
\providecommand \bibfield  [0]{\@secondoftwo}%
\providecommand \translation [1]{[#1]}%
\providecommand \BibitemOpen [0]{}%
\providecommand \bibitemStop [0]{}%
\providecommand \bibitemNoStop [0]{.\EOS\space}%
\providecommand \EOS [0]{\spacefactor3000\relax}%
\providecommand \BibitemShut  [1]{\csname bibitem#1\endcsname}%
\let\auto@bib@innerbib\@empty
\bibitem [{\citenamefont {Fitch}\ and\ \citenamefont
  {Tarbutt}(2021)}]{Fitch2021}%
  \BibitemOpen
  \bibfield  {author} {\bibinfo {author} {\bibfnamefont {N.}~\bibnamefont
  {Fitch}}\ and\ \bibinfo {author} {\bibfnamefont {M.}~\bibnamefont
  {Tarbutt}},\ }\bibfield  {title} {\bibinfo {title} {Laser-cooled molecules},\
  }\href {https://doi.org/https://doi.org/10.1016/bs.aamop.2021.04.003}
  {\bibfield  {journal} {\bibinfo  {journal} {Adv. At. Mol. Opt.}\ }\textbf
  {\bibinfo {volume} {70}},\ \bibinfo {pages} {157} (\bibinfo {year}
  {2021})}\BibitemShut {NoStop}%
\bibitem [{\citenamefont {{Di Rosa}}(2004)}]{DiRosa2004}%
  \BibitemOpen
  \bibfield  {author} {\bibinfo {author} {\bibfnamefont {M.~D.}\ \bibnamefont
  {{Di Rosa}}},\ }\bibfield  {title} {\bibinfo {title} {{Laser-cooling
  molecules}},\ }\href {https://doi.org/10.1140/epjd/e2004-00167-2} {\bibfield
  {journal} {\bibinfo  {journal} {Eur. Phys. J. D}\ }\textbf {\bibinfo {volume}
  {31}},\ \bibinfo {pages} {395} (\bibinfo {year} {2004})}\BibitemShut
  {NoStop}%
\bibitem [{\citenamefont {Stuhl}\ \emph {et~al.}(2008)\citenamefont {Stuhl},
  \citenamefont {Sawyer}, \citenamefont {Wang},\ and\ \citenamefont
  {Ye}}]{Stuhl2008}%
  \BibitemOpen
  \bibfield  {author} {\bibinfo {author} {\bibfnamefont {B.~K.}\ \bibnamefont
  {Stuhl}}, \bibinfo {author} {\bibfnamefont {B.~C.}\ \bibnamefont {Sawyer}},
  \bibinfo {author} {\bibfnamefont {D.}~\bibnamefont {Wang}},\ and\ \bibinfo
  {author} {\bibfnamefont {J.}~\bibnamefont {Ye}},\ }\bibfield  {title}
  {\bibinfo {title} {Magneto-optical trap for polar molecules},\ }\href
  {https://doi.org/10.1103/PhysRevLett.101.243002} {\bibfield  {journal}
  {\bibinfo  {journal} {Phys. Rev. Lett.}\ }\textbf {\bibinfo {volume} {101}},\
  \bibinfo {pages} {243002} (\bibinfo {year} {2008})}\BibitemShut {NoStop}%
\bibitem [{\citenamefont {Isaev}\ and\ \citenamefont
  {Berger}(2016)}]{Isaev2016}%
  \BibitemOpen
  \bibfield  {author} {\bibinfo {author} {\bibfnamefont {T.~A.}\ \bibnamefont
  {Isaev}}\ and\ \bibinfo {author} {\bibfnamefont {R.}~\bibnamefont {Berger}},\
  }\bibfield  {title} {\bibinfo {title} {Polyatomic candidates for cooling of
  molecules with lasers from simple theoretical concepts},\ }\href
  {https://doi.org/10.1103/PhysRevLett.116.063006} {\bibfield  {journal}
  {\bibinfo  {journal} {Phys. Rev. Lett.}\ }\textbf {\bibinfo {volume} {116}},\
  \bibinfo {pages} {063006} (\bibinfo {year} {2016})}\BibitemShut {NoStop}%
\bibitem [{\citenamefont {Barry}\ \emph {et~al.}(2014)\citenamefont {Barry},
  \citenamefont {McCarron}, \citenamefont {Norrgard}, \citenamefont
  {Steinecker},\ and\ \citenamefont {DeMille}}]{Barry2014}%
  \BibitemOpen
  \bibfield  {author} {\bibinfo {author} {\bibfnamefont {J.~F.}\ \bibnamefont
  {Barry}}, \bibinfo {author} {\bibfnamefont {D.~J.}\ \bibnamefont {McCarron}},
  \bibinfo {author} {\bibfnamefont {E.~B.}\ \bibnamefont {Norrgard}}, \bibinfo
  {author} {\bibfnamefont {M.~H.}\ \bibnamefont {Steinecker}},\ and\ \bibinfo
  {author} {\bibfnamefont {D.}~\bibnamefont {DeMille}},\ }\bibfield  {title}
  {\bibinfo {title} {{Magneto-optical trapping of a diatomic molecule}},\
  }\href {https://doi.org/10.1038/nature13634} {\bibfield  {journal} {\bibinfo
  {journal} {Nature}\ }\textbf {\bibinfo {volume} {512}},\ \bibinfo {pages}
  {286} (\bibinfo {year} {2014})}\BibitemShut {NoStop}%
\bibitem [{\citenamefont {Truppe}\ \emph {et~al.}(2017)\citenamefont {Truppe},
  \citenamefont {Williams}, \citenamefont {Hambach}, \citenamefont {Caldwell},
  \citenamefont {Fitch}, \citenamefont {Hinds}, \citenamefont {Sauer},\ and\
  \citenamefont {Tarbutt}}]{Truppe2017}%
  \BibitemOpen
  \bibfield  {author} {\bibinfo {author} {\bibfnamefont {S.}~\bibnamefont
  {Truppe}}, \bibinfo {author} {\bibfnamefont {H.~J.}\ \bibnamefont
  {Williams}}, \bibinfo {author} {\bibfnamefont {M.}~\bibnamefont {Hambach}},
  \bibinfo {author} {\bibfnamefont {L.}~\bibnamefont {Caldwell}}, \bibinfo
  {author} {\bibfnamefont {N.~J.}\ \bibnamefont {Fitch}}, \bibinfo {author}
  {\bibfnamefont {E.~A.}\ \bibnamefont {Hinds}}, \bibinfo {author}
  {\bibfnamefont {B.~E.}\ \bibnamefont {Sauer}},\ and\ \bibinfo {author}
  {\bibfnamefont {M.~R.}\ \bibnamefont {Tarbutt}},\ }\bibfield  {title}
  {\bibinfo {title} {{Molecules cooled below the Doppler limit}},\ }\href
  {https://doi.org/10.1038/nphys4241} {\bibfield  {journal} {\bibinfo
  {journal} {Nature Physics}\ }\textbf {\bibinfo {volume} {13}},\ \bibinfo
  {pages} {1173} (\bibinfo {year} {2017})}\BibitemShut {NoStop}%
\bibitem [{\citenamefont {Anderegg}\ \emph {et~al.}(2017)\citenamefont
  {Anderegg}, \citenamefont {Augenbraun}, \citenamefont {Chae}, \citenamefont
  {Hemmerling}, \citenamefont {Hutzler}, \citenamefont {Ravi}, \citenamefont
  {Collopy}, \citenamefont {Ye}, \citenamefont {Ketterle},\ and\ \citenamefont
  {Doyle}}]{Anderegg2017}%
  \BibitemOpen
  \bibfield  {author} {\bibinfo {author} {\bibfnamefont {L.}~\bibnamefont
  {Anderegg}}, \bibinfo {author} {\bibfnamefont {B.~L.}\ \bibnamefont
  {Augenbraun}}, \bibinfo {author} {\bibfnamefont {E.}~\bibnamefont {Chae}},
  \bibinfo {author} {\bibfnamefont {B.}~\bibnamefont {Hemmerling}}, \bibinfo
  {author} {\bibfnamefont {N.~R.}\ \bibnamefont {Hutzler}}, \bibinfo {author}
  {\bibfnamefont {A.}~\bibnamefont {Ravi}}, \bibinfo {author} {\bibfnamefont
  {A.}~\bibnamefont {Collopy}}, \bibinfo {author} {\bibfnamefont
  {J.}~\bibnamefont {Ye}}, \bibinfo {author} {\bibfnamefont {W.}~\bibnamefont
  {Ketterle}},\ and\ \bibinfo {author} {\bibfnamefont {J.~M.}\ \bibnamefont
  {Doyle}},\ }\bibfield  {title} {\bibinfo {title} {{Radio Frequency
  Magneto-Optical Trapping of CaF with High Density}},\ }\href
  {https://doi.org/10.1103/PhysRevLett.119.103201} {\bibfield  {journal}
  {\bibinfo  {journal} {Phys. Rev. Lett.}\ }\textbf {\bibinfo {volume} {119}},\
  \bibinfo {pages} {103201} (\bibinfo {year} {2017})}\BibitemShut {NoStop}%
\bibitem [{\citenamefont {Collopy}\ \emph {et~al.}(2018)\citenamefont
  {Collopy}, \citenamefont {Ding}, \citenamefont {Wu}, \citenamefont
  {Finneran}, \citenamefont {Anderegg}, \citenamefont {Augenbraun},
  \citenamefont {Doyle},\ and\ \citenamefont {Ye}}]{Collopy2018}%
  \BibitemOpen
  \bibfield  {author} {\bibinfo {author} {\bibfnamefont {A.~L.}\ \bibnamefont
  {Collopy}}, \bibinfo {author} {\bibfnamefont {S.}~\bibnamefont {Ding}},
  \bibinfo {author} {\bibfnamefont {Y.}~\bibnamefont {Wu}}, \bibinfo {author}
  {\bibfnamefont {I.~A.}\ \bibnamefont {Finneran}}, \bibinfo {author}
  {\bibfnamefont {L.}~\bibnamefont {Anderegg}}, \bibinfo {author}
  {\bibfnamefont {B.~L.}\ \bibnamefont {Augenbraun}}, \bibinfo {author}
  {\bibfnamefont {J.~M.}\ \bibnamefont {Doyle}},\ and\ \bibinfo {author}
  {\bibfnamefont {J.}~\bibnamefont {Ye}},\ }\bibfield  {title} {\bibinfo
  {title} {{3D Magneto-Optical Trap of Yttrium Monoxide}},\ }\href
  {https://doi.org/10.1103/PhysRevLett.121.213201} {\bibfield  {journal}
  {\bibinfo  {journal} {Phys. Rev. Lett.}\ }\textbf {\bibinfo {volume} {121}},\
  \bibinfo {pages} {213201} (\bibinfo {year} {2018})}\BibitemShut {NoStop}%
\bibitem [{\citenamefont {Baum}\ \emph {et~al.}(2020)\citenamefont {Baum},
  \citenamefont {Vilas}, \citenamefont {Hallas}, \citenamefont {Augenbraun},
  \citenamefont {Raval}, \citenamefont {Mitra},\ and\ \citenamefont
  {Doyle}}]{Baum2020}%
  \BibitemOpen
  \bibfield  {author} {\bibinfo {author} {\bibfnamefont {L.}~\bibnamefont
  {Baum}}, \bibinfo {author} {\bibfnamefont {N.~B.}\ \bibnamefont {Vilas}},
  \bibinfo {author} {\bibfnamefont {C.}~\bibnamefont {Hallas}}, \bibinfo
  {author} {\bibfnamefont {B.~L.}\ \bibnamefont {Augenbraun}}, \bibinfo
  {author} {\bibfnamefont {S.}~\bibnamefont {Raval}}, \bibinfo {author}
  {\bibfnamefont {D.}~\bibnamefont {Mitra}},\ and\ \bibinfo {author}
  {\bibfnamefont {J.~M.}\ \bibnamefont {Doyle}},\ }\bibfield  {title} {\bibinfo
  {title} {1d magneto-optical trap of polyatomic molecules},\ }\href
  {https://doi.org/10.1103/PhysRevLett.124.133201} {\bibfield  {journal}
  {\bibinfo  {journal} {Phys. Rev. Lett.}\ }\textbf {\bibinfo {volume} {124}},\
  \bibinfo {pages} {133201} (\bibinfo {year} {2020})}\BibitemShut {NoStop}%
\bibitem [{\citenamefont {Vilas}\ \emph {et~al.}(2022)\citenamefont {Vilas},
  \citenamefont {Hallas}, \citenamefont {Anderegg}, \citenamefont {Robichaud},
  \citenamefont {Winnicki}, \citenamefont {Mitra},\ and\ \citenamefont
  {Doyle}}]{Vilas2021}%
  \BibitemOpen
  \bibfield  {author} {\bibinfo {author} {\bibfnamefont {N.~B.}\ \bibnamefont
  {Vilas}}, \bibinfo {author} {\bibfnamefont {C.}~\bibnamefont {Hallas}},
  \bibinfo {author} {\bibfnamefont {L.}~\bibnamefont {Anderegg}}, \bibinfo
  {author} {\bibfnamefont {P.}~\bibnamefont {Robichaud}}, \bibinfo {author}
  {\bibfnamefont {A.}~\bibnamefont {Winnicki}}, \bibinfo {author}
  {\bibfnamefont {D.}~\bibnamefont {Mitra}},\ and\ \bibinfo {author}
  {\bibfnamefont {J.~M.}\ \bibnamefont {Doyle}},\ }\bibfield  {title} {\bibinfo
  {title} {Magneto-optical trapping and sub-doppler cooling of a polyatomic
  molecule},\ }\href {https://doi.org/10.1038/s41586-022-04620-5} {\bibfield
  {journal} {\bibinfo  {journal} {Nature}\ }\textbf {\bibinfo {volume} {606}},\
  \bibinfo {pages} {70} (\bibinfo {year} {2022})}\BibitemShut {NoStop}%
\bibitem [{\citenamefont {Tarbutt}\ \emph {et~al.}(2013)\citenamefont
  {Tarbutt}, \citenamefont {Sauer}, \citenamefont {Hudson},\ and\ \citenamefont
  {Hinds}}]{Tarbutt2013}%
  \BibitemOpen
  \bibfield  {author} {\bibinfo {author} {\bibfnamefont {M.~R.}\ \bibnamefont
  {Tarbutt}}, \bibinfo {author} {\bibfnamefont {B.~E.}\ \bibnamefont {Sauer}},
  \bibinfo {author} {\bibfnamefont {J.~J.}\ \bibnamefont {Hudson}},\ and\
  \bibinfo {author} {\bibfnamefont {E.~A.}\ \bibnamefont {Hinds}},\ }\bibfield
  {title} {\bibinfo {title} {Design for a fountain of {YbF} molecules to
  measure the electron{\textquotesingle}s electric dipole moment},\ }\href
  {https://doi.org/10.1088/1367-2630/15/5/053034} {\bibfield  {journal}
  {\bibinfo  {journal} {New Journal of Physics}\ }\textbf {\bibinfo {volume}
  {15}},\ \bibinfo {pages} {053034} (\bibinfo {year} {2013})}\BibitemShut
  {NoStop}%
\bibitem [{\citenamefont {Kozyryev}\ and\ \citenamefont
  {Hutzler}(2017)}]{Kozyryev2017}%
  \BibitemOpen
  \bibfield  {author} {\bibinfo {author} {\bibfnamefont {I.}~\bibnamefont
  {Kozyryev}}\ and\ \bibinfo {author} {\bibfnamefont {N.~R.}\ \bibnamefont
  {Hutzler}},\ }\bibfield  {title} {\bibinfo {title} {Precision measurement of
  time-reversal symmetry violation with laser-cooled polyatomic molecules},\
  }\href {https://doi.org/10.1103/PhysRevLett.119.133002} {\bibfield  {journal}
  {\bibinfo  {journal} {Phys. Rev. Lett.}\ }\textbf {\bibinfo {volume} {119}},\
  \bibinfo {pages} {133002} (\bibinfo {year} {2017})}\BibitemShut {NoStop}%
\bibitem [{\citenamefont {Norrgard}\ \emph {et~al.}(2017)\citenamefont
  {Norrgard}, \citenamefont {Edwards}, \citenamefont {McCarron}, \citenamefont
  {Steinecker}, \citenamefont {DeMille}, \citenamefont {Alam}, \citenamefont
  {Peck}, \citenamefont {Wadia},\ and\ \citenamefont {Hunter}}]{Norrgard2017}%
  \BibitemOpen
  \bibfield  {author} {\bibinfo {author} {\bibfnamefont {E.~B.}\ \bibnamefont
  {Norrgard}}, \bibinfo {author} {\bibfnamefont {E.~R.}\ \bibnamefont
  {Edwards}}, \bibinfo {author} {\bibfnamefont {D.~J.}\ \bibnamefont
  {McCarron}}, \bibinfo {author} {\bibfnamefont {M.~H.}\ \bibnamefont
  {Steinecker}}, \bibinfo {author} {\bibfnamefont {D.}~\bibnamefont {DeMille}},
  \bibinfo {author} {\bibfnamefont {S.~S.}\ \bibnamefont {Alam}}, \bibinfo
  {author} {\bibfnamefont {S.~K.}\ \bibnamefont {Peck}}, \bibinfo {author}
  {\bibfnamefont {N.~S.}\ \bibnamefont {Wadia}},\ and\ \bibinfo {author}
  {\bibfnamefont {L.~R.}\ \bibnamefont {Hunter}},\ }\bibfield  {title}
  {\bibinfo {title} {Hyperfine structure of the
  {$B{\phantom{\rule{0.16em}{0ex}}}^{3}{\mathrm{\ensuremath{\Pi}}}_{1}$} state
  and predictions of optical cycling behavior in the
  {$X\ensuremath{\rightarrow}B$} transition of {TlF}},\ }\href
  {https://doi.org/10.1103/PhysRevA.95.062506} {\bibfield  {journal} {\bibinfo
  {journal} {Phys. Rev. A}\ }\textbf {\bibinfo {volume} {95}},\ \bibinfo
  {pages} {062506} (\bibinfo {year} {2017})}\BibitemShut {NoStop}%
\bibitem [{\citenamefont {Aggarwal}\ \emph {et~al.}(2018)\citenamefont
  {Aggarwal}, \citenamefont {Bethlem}, \citenamefont {Borschevsky},
  \citenamefont {Denis}, \citenamefont {Esajas}, \citenamefont {Haase},
  \citenamefont {Hao}, \citenamefont {Hoekstra}, \citenamefont {Jungmann},
  \citenamefont {Meijknecht}, \citenamefont {Mooij}, \citenamefont
  {Timmermans}, \citenamefont {Ubachs}, \citenamefont {Willmann},\ and\
  \citenamefont {Zapara}}]{Aggarwal2018}%
  \BibitemOpen
  \bibfield  {author} {\bibinfo {author} {\bibfnamefont {P.}~\bibnamefont
  {Aggarwal}}, \bibinfo {author} {\bibfnamefont {H.~L.}\ \bibnamefont
  {Bethlem}}, \bibinfo {author} {\bibfnamefont {A.}~\bibnamefont
  {Borschevsky}}, \bibinfo {author} {\bibfnamefont {M.}~\bibnamefont {Denis}},
  \bibinfo {author} {\bibfnamefont {K.}~\bibnamefont {Esajas}}, \bibinfo
  {author} {\bibfnamefont {P.~A.~B.}\ \bibnamefont {Haase}}, \bibinfo {author}
  {\bibfnamefont {Y.}~\bibnamefont {Hao}}, \bibinfo {author} {\bibfnamefont
  {S.}~\bibnamefont {Hoekstra}}, \bibinfo {author} {\bibfnamefont
  {K.}~\bibnamefont {Jungmann}}, \bibinfo {author} {\bibfnamefont {T.~B.}\
  \bibnamefont {Meijknecht}}, \bibinfo {author} {\bibfnamefont {M.~C.}\
  \bibnamefont {Mooij}}, \bibinfo {author} {\bibfnamefont {R.~G.~E.}\
  \bibnamefont {Timmermans}}, \bibinfo {author} {\bibfnamefont
  {W.}~\bibnamefont {Ubachs}}, \bibinfo {author} {\bibfnamefont
  {L.}~\bibnamefont {Willmann}},\ and\ \bibinfo {author} {\bibfnamefont
  {A.}~\bibnamefont {Zapara}},\ }\bibfield  {title} {\bibinfo {title}
  {{Measuring the electric dipole moment of the electron in BaF}},\ }\href
  {https://doi.org/10.1140/epjd/e2018-90192-9} {\bibfield  {journal} {\bibinfo
  {journal} {Eur. Phys. J. D}\ }\textbf {\bibinfo {volume} {72}},\ \bibinfo
  {pages} {197} (\bibinfo {year} {2018})}\BibitemShut {NoStop}%
\bibitem [{\citenamefont {Altuntas}\ \emph {et~al.}(2018)\citenamefont
  {Altuntas}, \citenamefont {Ammon}, \citenamefont {Cahn},\ and\ \citenamefont
  {DeMille}}]{Altuntas2018}%
  \BibitemOpen
  \bibfield  {author} {\bibinfo {author} {\bibfnamefont {E.}~\bibnamefont
  {Altuntas}}, \bibinfo {author} {\bibfnamefont {J.}~\bibnamefont {Ammon}},
  \bibinfo {author} {\bibfnamefont {S.~B.}\ \bibnamefont {Cahn}},\ and\
  \bibinfo {author} {\bibfnamefont {D.}~\bibnamefont {DeMille}},\ }\bibfield
  {title} {\bibinfo {title} {{Demonstration of a Sensitive Method to Measure
  Nuclear-Spin-Dependent Parity Violation}},\ }\href
  {https://doi.org/10.1103/PhysRevLett.120.142501} {\bibfield  {journal}
  {\bibinfo  {journal} {Phys. Rev. Lett.}\ }\textbf {\bibinfo {volume} {120}},\
  \bibinfo {pages} {142501} (\bibinfo {year} {2018})}\BibitemShut {NoStop}%
\bibitem [{\citenamefont {Lim}\ \emph {et~al.}(2018)\citenamefont {Lim},
  \citenamefont {Almond}, \citenamefont {Trigatzis}, \citenamefont {Devlin},
  \citenamefont {Fitch}, \citenamefont {Sauer}, \citenamefont {Tarbutt},\ and\
  \citenamefont {Hinds}}]{Lim2018}%
  \BibitemOpen
  \bibfield  {author} {\bibinfo {author} {\bibfnamefont {J.}~\bibnamefont
  {Lim}}, \bibinfo {author} {\bibfnamefont {J.~R.}\ \bibnamefont {Almond}},
  \bibinfo {author} {\bibfnamefont {M.~A.}\ \bibnamefont {Trigatzis}}, \bibinfo
  {author} {\bibfnamefont {J.~A.}\ \bibnamefont {Devlin}}, \bibinfo {author}
  {\bibfnamefont {N.~J.}\ \bibnamefont {Fitch}}, \bibinfo {author}
  {\bibfnamefont {B.~E.}\ \bibnamefont {Sauer}}, \bibinfo {author}
  {\bibfnamefont {M.~R.}\ \bibnamefont {Tarbutt}},\ and\ \bibinfo {author}
  {\bibfnamefont {E.~A.}\ \bibnamefont {Hinds}},\ }\bibfield  {title} {\bibinfo
  {title} {Laser cooled {YbF} molecules for measuring the electron's electric
  dipole moment},\ }\href {https://doi.org/10.1103/PhysRevLett.120.123201}
  {\bibfield  {journal} {\bibinfo  {journal} {Phys. Rev. Lett.}\ }\textbf
  {\bibinfo {volume} {120}},\ \bibinfo {pages} {123201} (\bibinfo {year}
  {2018})}\BibitemShut {NoStop}%
\bibitem [{\citenamefont {O'Rourke}\ and\ \citenamefont
  {Hutzler}(2019)}]{Rourke2019}%
  \BibitemOpen
  \bibfield  {author} {\bibinfo {author} {\bibfnamefont {M.~J.}\ \bibnamefont
  {O'Rourke}}\ and\ \bibinfo {author} {\bibfnamefont {N.~R.}\ \bibnamefont
  {Hutzler}},\ }\bibfield  {title} {\bibinfo {title} {Hypermetallic polar
  molecules for precision measurements},\ }\href
  {https://doi.org/10.1103/PhysRevA.100.022502} {\bibfield  {journal} {\bibinfo
   {journal} {Phys. Rev. A}\ }\textbf {\bibinfo {volume} {100}},\ \bibinfo
  {pages} {022502} (\bibinfo {year} {2019})}\BibitemShut {NoStop}%
\bibitem [{\citenamefont {{Garcia Ruiz \textit{et
  al.}}}(2020)}]{GarciaRuiz2020}%
  \BibitemOpen
  \bibfield  {author} {\bibinfo {author} {\bibfnamefont {R.}~\bibnamefont
  {{Garcia Ruiz \textit{et al.}}}},\ }\bibfield  {title} {\bibinfo {title}
  {{Spectroscopy of short-lived radioactive molecules}},\ }\href
  {https://doi.org/10.1038/s41586-020-2299-4} {\bibfield  {journal} {\bibinfo
  {journal} {Nature}\ }\textbf {\bibinfo {volume} {581}},\ \bibinfo {pages}
  {396} (\bibinfo {year} {2020})}\BibitemShut {NoStop}%
\bibitem [{\citenamefont {McNally}\ \emph {et~al.}(2020)\citenamefont
  {McNally}, \citenamefont {Kozyryev}, \citenamefont {Vazquez-Carson},
  \citenamefont {Wenz}, \citenamefont {Wang},\ and\ \citenamefont
  {Zelevinsky}}]{McNally2020}%
  \BibitemOpen
  \bibfield  {author} {\bibinfo {author} {\bibfnamefont {R.~L.}\ \bibnamefont
  {McNally}}, \bibinfo {author} {\bibfnamefont {I.}~\bibnamefont {Kozyryev}},
  \bibinfo {author} {\bibfnamefont {S.}~\bibnamefont {Vazquez-Carson}},
  \bibinfo {author} {\bibfnamefont {K.}~\bibnamefont {Wenz}}, \bibinfo {author}
  {\bibfnamefont {T.}~\bibnamefont {Wang}},\ and\ \bibinfo {author}
  {\bibfnamefont {T.}~\bibnamefont {Zelevinsky}},\ }\bibfield  {title}
  {\bibinfo {title} {Optical cycling, radiative deflection and laser cooling of
  barium monohydride (${}^{138}${BaH})},\ }\href
  {https://doi.org/10.1088/1367-2630/aba3e9} {\bibfield  {journal} {\bibinfo
  {journal} {New Journal of Physics}\ }\textbf {\bibinfo {volume} {22}},\
  \bibinfo {pages} {083047} (\bibinfo {year} {2020})}\BibitemShut {NoStop}%
\bibitem [{\citenamefont {Mitra}\ \emph {et~al.}(2020)\citenamefont {Mitra},
  \citenamefont {Vilas}, \citenamefont {Hallas}, \citenamefont {Anderegg},
  \citenamefont {Augenbraun}, \citenamefont {Baum}, \citenamefont {Miller},
  \citenamefont {Raval},\ and\ \citenamefont {Doyle}}]{Mitra2020}%
  \BibitemOpen
  \bibfield  {author} {\bibinfo {author} {\bibfnamefont {D.}~\bibnamefont
  {Mitra}}, \bibinfo {author} {\bibfnamefont {N.~B.}\ \bibnamefont {Vilas}},
  \bibinfo {author} {\bibfnamefont {C.}~\bibnamefont {Hallas}}, \bibinfo
  {author} {\bibfnamefont {L.}~\bibnamefont {Anderegg}}, \bibinfo {author}
  {\bibfnamefont {B.~L.}\ \bibnamefont {Augenbraun}}, \bibinfo {author}
  {\bibfnamefont {L.}~\bibnamefont {Baum}}, \bibinfo {author} {\bibfnamefont
  {C.}~\bibnamefont {Miller}}, \bibinfo {author} {\bibfnamefont
  {S.}~\bibnamefont {Raval}},\ and\ \bibinfo {author} {\bibfnamefont {J.~M.}\
  \bibnamefont {Doyle}},\ }\bibfield  {title} {\bibinfo {title} {Direct laser
  cooling of a symmetric top molecule},\ }\href
  {https://doi.org/10.1126/science.abc5357} {\bibfield  {journal} {\bibinfo
  {journal} {Science}\ }\textbf {\bibinfo {volume} {369}},\ \bibinfo {pages}
  {1366} (\bibinfo {year} {2020})}\BibitemShut {NoStop}%
\bibitem [{\citenamefont {K\l{}os}\ and\ \citenamefont
  {Kotochigova}(2020)}]{Klos2020}%
  \BibitemOpen
  \bibfield  {author} {\bibinfo {author} {\bibfnamefont {J.}~\bibnamefont
  {K\l{}os}}\ and\ \bibinfo {author} {\bibfnamefont {S.}~\bibnamefont
  {Kotochigova}},\ }\bibfield  {title} {\bibinfo {title} {Prospects for laser
  cooling of polyatomic molecules with increasing complexity},\ }\href
  {https://doi.org/10.1103/PhysRevResearch.2.013384} {\bibfield  {journal}
  {\bibinfo  {journal} {Phys. Rev. Research}\ }\textbf {\bibinfo {volume}
  {2}},\ \bibinfo {pages} {013384} (\bibinfo {year} {2020})}\BibitemShut
  {NoStop}%
\bibitem [{\citenamefont {Cheuk}\ \emph {et~al.}(2020)\citenamefont {Cheuk},
  \citenamefont {Anderegg}, \citenamefont {Bao}, \citenamefont {Burchesky},
  \citenamefont {Yu}, \citenamefont {Ketterle}, \citenamefont {Ni},\ and\
  \citenamefont {Doyle}}]{Cheuk2020}%
  \BibitemOpen
  \bibfield  {author} {\bibinfo {author} {\bibfnamefont {L.~W.}\ \bibnamefont
  {Cheuk}}, \bibinfo {author} {\bibfnamefont {L.}~\bibnamefont {Anderegg}},
  \bibinfo {author} {\bibfnamefont {Y.}~\bibnamefont {Bao}}, \bibinfo {author}
  {\bibfnamefont {S.}~\bibnamefont {Burchesky}}, \bibinfo {author}
  {\bibfnamefont {S.~S.}\ \bibnamefont {Yu}}, \bibinfo {author} {\bibfnamefont
  {W.}~\bibnamefont {Ketterle}}, \bibinfo {author} {\bibfnamefont {K.-K.}\
  \bibnamefont {Ni}},\ and\ \bibinfo {author} {\bibfnamefont {J.~M.}\
  \bibnamefont {Doyle}},\ }\bibfield  {title} {\bibinfo {title} {{Observation
  of Collisions between Two Ultracold Ground-State CaF Molecules}},\ }\href
  {https://doi.org/10.1103/PhysRevLett.125.043401} {\bibfield  {journal}
  {\bibinfo  {journal} {Phys. Rev. Lett.}\ }\textbf {\bibinfo {volume} {125}},\
  \bibinfo {pages} {43401} (\bibinfo {year} {2020})}\BibitemShut {NoStop}%
\bibitem [{\citenamefont {Blackmore}\ \emph {et~al.}(2018)\citenamefont
  {Blackmore}, \citenamefont {Caldwell}, \citenamefont {Gregory}, \citenamefont
  {Bridge}, \citenamefont {Sawant}, \citenamefont {Aldegunde}, \citenamefont
  {Mur-Petit}, \citenamefont {Jaksch}, \citenamefont {Hutson}, \citenamefont
  {Sauer}, \citenamefont {Tarbutt},\ and\ \citenamefont
  {Cornish}}]{Blackmore2018}%
  \BibitemOpen
  \bibfield  {author} {\bibinfo {author} {\bibfnamefont {J.~A.}\ \bibnamefont
  {Blackmore}}, \bibinfo {author} {\bibfnamefont {L.}~\bibnamefont {Caldwell}},
  \bibinfo {author} {\bibfnamefont {P.~D.}\ \bibnamefont {Gregory}}, \bibinfo
  {author} {\bibfnamefont {E.~M.}\ \bibnamefont {Bridge}}, \bibinfo {author}
  {\bibfnamefont {R.}~\bibnamefont {Sawant}}, \bibinfo {author} {\bibfnamefont
  {J.}~\bibnamefont {Aldegunde}}, \bibinfo {author} {\bibfnamefont
  {J.}~\bibnamefont {Mur-Petit}}, \bibinfo {author} {\bibfnamefont
  {D.}~\bibnamefont {Jaksch}}, \bibinfo {author} {\bibfnamefont {J.~M.}\
  \bibnamefont {Hutson}}, \bibinfo {author} {\bibfnamefont {B.~E.}\
  \bibnamefont {Sauer}}, \bibinfo {author} {\bibfnamefont {M.~R.}\ \bibnamefont
  {Tarbutt}},\ and\ \bibinfo {author} {\bibfnamefont {S.~L.}\ \bibnamefont
  {Cornish}},\ }\bibfield  {title} {\bibinfo {title} {Ultracold molecules for
  quantum simulation: rotational coherences in {CaF} and {RbCs}},\ }\href
  {https://doi.org/10.1088/2058-9565/aaee35} {\bibfield  {journal} {\bibinfo
  {journal} {Quantum Science and Technology}\ }\textbf {\bibinfo {volume}
  {4}},\ \bibinfo {pages} {014010} (\bibinfo {year} {2018})}\BibitemShut
  {NoStop}%
\bibitem [{\citenamefont {Anderegg}\ \emph {et~al.}(2019)\citenamefont
  {Anderegg}, \citenamefont {Cheuk}, \citenamefont {Bao}, \citenamefont
  {Burchesky}, \citenamefont {Ketterle}, \citenamefont {Ni},\ and\
  \citenamefont {Doyle}}]{Anderegg2019}%
  \BibitemOpen
  \bibfield  {author} {\bibinfo {author} {\bibfnamefont {L.}~\bibnamefont
  {Anderegg}}, \bibinfo {author} {\bibfnamefont {L.~W.}\ \bibnamefont {Cheuk}},
  \bibinfo {author} {\bibfnamefont {Y.}~\bibnamefont {Bao}}, \bibinfo {author}
  {\bibfnamefont {S.}~\bibnamefont {Burchesky}}, \bibinfo {author}
  {\bibfnamefont {W.}~\bibnamefont {Ketterle}}, \bibinfo {author}
  {\bibfnamefont {K.-K.}\ \bibnamefont {Ni}},\ and\ \bibinfo {author}
  {\bibfnamefont {J.~M.}\ \bibnamefont {Doyle}},\ }\bibfield  {title} {\bibinfo
  {title} {{An optical tweezer array of ultracold molecules}},\ }\href
  {https://doi.org/10.1126/science.aax1265} {\bibfield  {journal} {\bibinfo
  {journal} {Science}\ }\textbf {\bibinfo {volume} {365}},\ \bibinfo {pages}
  {1156} (\bibinfo {year} {2019})}\BibitemShut {NoStop}%
\bibitem [{\citenamefont {Caldwell}\ and\ \citenamefont
  {Tarbutt}(2020)}]{Caldwell2020}%
  \BibitemOpen
  \bibfield  {author} {\bibinfo {author} {\bibfnamefont {L.}~\bibnamefont
  {Caldwell}}\ and\ \bibinfo {author} {\bibfnamefont {M.~R.}\ \bibnamefont
  {Tarbutt}},\ }\bibfield  {title} {\bibinfo {title} {Enhancing dipolar
  interactions between molecules using state-dependent optical tweezer traps},\
  }\href {https://doi.org/10.1103/PhysRevLett.125.243201} {\bibfield  {journal}
  {\bibinfo  {journal} {Phys. Rev. Lett.}\ }\textbf {\bibinfo {volume} {125}},\
  \bibinfo {pages} {243201} (\bibinfo {year} {2020})}\BibitemShut {NoStop}%
\bibitem [{\citenamefont {Chae}(2021)}]{Chae2021}%
  \BibitemOpen
  \bibfield  {author} {\bibinfo {author} {\bibfnamefont {E.}~\bibnamefont
  {Chae}},\ }\bibfield  {title} {\bibinfo {title} {Entanglement via rotational
  blockade of mgf molecules in a magic potential},\ }\href
  {https://doi.org/10.1039/D0CP04042H} {\bibfield  {journal} {\bibinfo
  {journal} {Phys. Chem. Chem. Phys.}\ }\textbf {\bibinfo {volume} {23}},\
  \bibinfo {pages} {1215} (\bibinfo {year} {2021})}\BibitemShut {NoStop}%
\bibitem [{\citenamefont {Chen}\ \emph {et~al.}(2017)\citenamefont {Chen},
  \citenamefont {Bu},\ and\ \citenamefont {Yan}}]{Chen2017}%
  \BibitemOpen
  \bibfield  {author} {\bibinfo {author} {\bibfnamefont {T.}~\bibnamefont
  {Chen}}, \bibinfo {author} {\bibfnamefont {W.}~\bibnamefont {Bu}},\ and\
  \bibinfo {author} {\bibfnamefont {B.}~\bibnamefont {Yan}},\ }\bibfield
  {title} {\bibinfo {title} {Radiative deflection of a {BaF} molecular beam via
  optical cycling},\ }\href {https://doi.org/10.1103/PhysRevA.96.053401}
  {\bibfield  {journal} {\bibinfo  {journal} {Phys. Rev. A}\ }\textbf {\bibinfo
  {volume} {96}},\ \bibinfo {pages} {053401} (\bibinfo {year}
  {2017})}\BibitemShut {NoStop}%
\bibitem [{\citenamefont {Albrecht}\ \emph {et~al.}(2020)\citenamefont
  {Albrecht}, \citenamefont {Scharwaechter}, \citenamefont {Sixt},
  \citenamefont {Hofer},\ and\ \citenamefont {Langen}}]{Albrecht2020}%
  \BibitemOpen
  \bibfield  {author} {\bibinfo {author} {\bibfnamefont {R.}~\bibnamefont
  {Albrecht}}, \bibinfo {author} {\bibfnamefont {M.}~\bibnamefont
  {Scharwaechter}}, \bibinfo {author} {\bibfnamefont {T.}~\bibnamefont {Sixt}},
  \bibinfo {author} {\bibfnamefont {L.}~\bibnamefont {Hofer}},\ and\ \bibinfo
  {author} {\bibfnamefont {T.}~\bibnamefont {Langen}},\ }\bibfield  {title}
  {\bibinfo {title} {Buffer-gas cooling, high-resolution spectroscopy, and
  optical cycling of barium monofluoride molecules},\ }\href
  {https://doi.org/10.1103/PhysRevA.101.013413} {\bibfield  {journal} {\bibinfo
   {journal} {Phys. Rev. A}\ }\textbf {\bibinfo {volume} {101}},\ \bibinfo
  {pages} {013413} (\bibinfo {year} {2020})}\BibitemShut {NoStop}%
\bibitem [{\citenamefont {DeMille}\ \emph {et~al.}(2008)\citenamefont
  {DeMille}, \citenamefont {Cahn}, \citenamefont {Murphree}, \citenamefont
  {Rahmlow},\ and\ \citenamefont {Kozlov}}]{Demille2008}%
  \BibitemOpen
  \bibfield  {author} {\bibinfo {author} {\bibfnamefont {D.}~\bibnamefont
  {DeMille}}, \bibinfo {author} {\bibfnamefont {S.~B.}\ \bibnamefont {Cahn}},
  \bibinfo {author} {\bibfnamefont {D.}~\bibnamefont {Murphree}}, \bibinfo
  {author} {\bibfnamefont {D.~A.}\ \bibnamefont {Rahmlow}},\ and\ \bibinfo
  {author} {\bibfnamefont {M.~G.}\ \bibnamefont {Kozlov}},\ }\bibfield  {title}
  {\bibinfo {title} {{Using Molecules to Measure Nuclear Spin-Dependent Parity
  Violation}},\ }\href {https://doi.org/10.1103/PhysRevLett.100.023003}
  {\bibfield  {journal} {\bibinfo  {journal} {Phys. Rev. Lett.}\ }\textbf
  {\bibinfo {volume} {100}},\ \bibinfo {pages} {23003} (\bibinfo {year}
  {2008})}\BibitemShut {NoStop}%
\bibitem [{\citenamefont {Kogel}\ \emph {et~al.}(2021)\citenamefont {Kogel},
  \citenamefont {Rockenhäuser}, \citenamefont {Albrecht},\ and\ \citenamefont
  {Langen}}]{Kogel2021}%
  \BibitemOpen
  \bibfield  {author} {\bibinfo {author} {\bibfnamefont {F.}~\bibnamefont
  {Kogel}}, \bibinfo {author} {\bibfnamefont {M.}~\bibnamefont
  {Rockenhäuser}}, \bibinfo {author} {\bibfnamefont {R.}~\bibnamefont
  {Albrecht}},\ and\ \bibinfo {author} {\bibfnamefont {T.}~\bibnamefont
  {Langen}},\ }\bibfield  {title} {\bibinfo {title} {A laser cooling scheme for
  precision measurements using fermionic barium monofluoride ({137Ba19F})
  molecules},\ }\href {https://doi.org/10.1088/1367-2630/ac1df2} {\bibfield
  {journal} {\bibinfo  {journal} {New Journal of Physics}\ }\textbf {\bibinfo
  {volume} {23}},\ \bibinfo {pages} {095003} (\bibinfo {year}
  {2021})}\BibitemShut {NoStop}%
\bibitem [{\citenamefont {Vutha}\ \emph {et~al.}(2018)\citenamefont {Vutha},
  \citenamefont {Horbatsch},\ and\ \citenamefont {Hessels}}]{Vutha2018}%
  \BibitemOpen
  \bibfield  {author} {\bibinfo {author} {\bibfnamefont {A.~C.}\ \bibnamefont
  {Vutha}}, \bibinfo {author} {\bibfnamefont {M.}~\bibnamefont {Horbatsch}},\
  and\ \bibinfo {author} {\bibfnamefont {E.~A.}\ \bibnamefont {Hessels}},\
  }\bibfield  {title} {\bibinfo {title} {Orientation-dependent hyperfine
  structure of polar molecules in a rare-gas matrix: A scheme for measuring the
  electron electric dipole moment},\ }\href
  {https://doi.org/10.1103/PhysRevA.98.032513} {\bibfield  {journal} {\bibinfo
  {journal} {Phys. Rev. A}\ }\textbf {\bibinfo {volume} {98}},\ \bibinfo
  {pages} {032513} (\bibinfo {year} {2018})}\BibitemShut {NoStop}%
\bibitem [{\citenamefont {Denis}\ \emph {et~al.}(2022)\citenamefont {Denis},
  \citenamefont {Haase}, \citenamefont {Mooij}, \citenamefont {Chamorro},
  \citenamefont {Aggarwal}, \citenamefont {Bethlem}, \citenamefont
  {Boeschoten}, \citenamefont {Borschevsky}, \citenamefont {Esajas},
  \citenamefont {Hao}, \citenamefont {Hoekstra}, \citenamefont {van Hofslot},
  \citenamefont {Marshall}, \citenamefont {Meijknecht}, \citenamefont
  {Timmermans}, \citenamefont {Touwen}, \citenamefont {Ubachs}, \citenamefont
  {Willmann},\ and\ \citenamefont {Yin}}]{Denis2022}%
  \BibitemOpen
  \bibfield  {author} {\bibinfo {author} {\bibfnamefont {M.}~\bibnamefont
  {Denis}}, \bibinfo {author} {\bibfnamefont {P.~A.~B.}\ \bibnamefont {Haase}},
  \bibinfo {author} {\bibfnamefont {M.~C.}\ \bibnamefont {Mooij}}, \bibinfo
  {author} {\bibfnamefont {Y.}~\bibnamefont {Chamorro}}, \bibinfo {author}
  {\bibfnamefont {P.}~\bibnamefont {Aggarwal}}, \bibinfo {author}
  {\bibfnamefont {H.~L.}\ \bibnamefont {Bethlem}}, \bibinfo {author}
  {\bibfnamefont {A.}~\bibnamefont {Boeschoten}}, \bibinfo {author}
  {\bibfnamefont {A.}~\bibnamefont {Borschevsky}}, \bibinfo {author}
  {\bibfnamefont {K.}~\bibnamefont {Esajas}}, \bibinfo {author} {\bibfnamefont
  {Y.}~\bibnamefont {Hao}}, \bibinfo {author} {\bibfnamefont {S.}~\bibnamefont
  {Hoekstra}}, \bibinfo {author} {\bibfnamefont {J.~W.~F.}\ \bibnamefont {van
  Hofslot}}, \bibinfo {author} {\bibfnamefont {V.~R.}\ \bibnamefont
  {Marshall}}, \bibinfo {author} {\bibfnamefont {T.~B.}\ \bibnamefont
  {Meijknecht}}, \bibinfo {author} {\bibfnamefont {R.~G.~E.}\ \bibnamefont
  {Timmermans}}, \bibinfo {author} {\bibfnamefont {A.}~\bibnamefont {Touwen}},
  \bibinfo {author} {\bibfnamefont {W.}~\bibnamefont {Ubachs}}, \bibinfo
  {author} {\bibfnamefont {L.}~\bibnamefont {Willmann}},\ and\ \bibinfo
  {author} {\bibfnamefont {Y.}~\bibnamefont {Yin}} (\bibinfo {collaboration}
  {NL-$e\text{EDM}$ Collaboration}),\ }\bibfield  {title} {\bibinfo {title}
  {Benchmarking of the fock-space coupled-cluster method and uncertainty
  estimation: Magnetic hyperfine interaction in the excited state of {BaF}},\
  }\href {https://doi.org/10.1103/PhysRevA.105.052811} {\bibfield  {journal}
  {\bibinfo  {journal} {Phys. Rev. A}\ }\textbf {\bibinfo {volume} {105}},\
  \bibinfo {pages} {052811} (\bibinfo {year} {2022})}\BibitemShut {NoStop}%
\bibitem [{\citenamefont {Marshall}()}]{Marshall2022}%
  \BibitemOpen
  \bibfield  {author} {\bibinfo {author} {\bibfnamefont {V.~R.}\ \bibnamefont
  {Marshall}},\ }\href@noop {} {\bibinfo {title} {{et al. (NL-eEDM
  collaboration)}, private communication}}\BibitemShut {NoStop}%
\bibitem [{\citenamefont {Bu}\ \emph {et~al.}(2022)\citenamefont {Bu},
  \citenamefont {Zhang}, \citenamefont {Liang}, \citenamefont {Chen},\ and\
  \citenamefont {Yan}}]{Bu2022}%
  \BibitemOpen
  \bibfield  {author} {\bibinfo {author} {\bibfnamefont {W.}~\bibnamefont
  {Bu}}, \bibinfo {author} {\bibfnamefont {Y.}~\bibnamefont {Zhang}}, \bibinfo
  {author} {\bibfnamefont {Q.}~\bibnamefont {Liang}}, \bibinfo {author}
  {\bibfnamefont {T.}~\bibnamefont {Chen}},\ and\ \bibinfo {author}
  {\bibfnamefont {B.}~\bibnamefont {Yan}},\ }\bibfield  {title} {\bibinfo
  {title} {{Saturated absorption spectroscopy of buffer-gas-cooled Barium
  monofluoride molecules}},\ }\href {https://doi.org/10.1007/s11467-022-1194-x}
  {\bibfield  {journal} {\bibinfo  {journal} {Frontiers of Physics}\ }\textbf
  {\bibinfo {volume} {17}},\ \bibinfo {pages} {62502} (\bibinfo {year}
  {2022})}\BibitemShut {NoStop}%
\bibitem [{\citenamefont {Ryzlewicz}\ and\ \citenamefont
  {Törring}(1980)}]{Ryzlewicz1980}%
  \BibitemOpen
  \bibfield  {author} {\bibinfo {author} {\bibfnamefont {C.}~\bibnamefont
  {Ryzlewicz}}\ and\ \bibinfo {author} {\bibfnamefont {T.}~\bibnamefont
  {Törring}},\ }\bibfield  {title} {\bibinfo {title} {Formation and microwave
  spectrum of the {${}^2\Sigma$} radical barium monofluoride},\ }\href
  {https://doi.org/https://doi.org/10.1016/0301-0104(80)80107-8} {\bibfield
  {journal} {\bibinfo  {journal} {Chemical Physics}\ }\textbf {\bibinfo
  {volume} {51}},\ \bibinfo {pages} {329} (\bibinfo {year} {1980})}\BibitemShut
  {NoStop}%
\bibitem [{\citenamefont {Ernst}\ \emph {et~al.}(1986)\citenamefont {Ernst},
  \citenamefont {K{\"{a}}ndler},\ and\ \citenamefont
  {T{\"{o}}rring}}]{Ernst1986}%
  \BibitemOpen
  \bibfield  {author} {\bibinfo {author} {\bibfnamefont {W.~E.}\ \bibnamefont
  {Ernst}}, \bibinfo {author} {\bibfnamefont {J.}~\bibnamefont
  {K{\"{a}}ndler}},\ and\ \bibinfo {author} {\bibfnamefont {T.}~\bibnamefont
  {T{\"{o}}rring}},\ }\bibfield  {title} {\bibinfo {title} {{Hyperfine
  structure and electric dipole moment of BaF X2$\Sigma$+}},\ }\href
  {https://doi.org/10.1063/1.449961} {\bibfield  {journal} {\bibinfo  {journal}
  {The Journal of Chemical Physics}\ }\textbf {\bibinfo {volume} {84}},\
  \bibinfo {pages} {4769} (\bibinfo {year} {1986})}\BibitemShut {NoStop}%
\bibitem [{\citenamefont {Effantin}\ \emph {et~al.}(1990)\citenamefont
  {Effantin}, \citenamefont {Bernard}, \citenamefont {d'Incan}, \citenamefont
  {Wannous}, \citenamefont {Vergès},\ and\ \citenamefont
  {Barrow}}]{Effantin1990}%
  \BibitemOpen
  \bibfield  {author} {\bibinfo {author} {\bibfnamefont {C.}~\bibnamefont
  {Effantin}}, \bibinfo {author} {\bibfnamefont {A.}~\bibnamefont {Bernard}},
  \bibinfo {author} {\bibfnamefont {J.}~\bibnamefont {d'Incan}}, \bibinfo
  {author} {\bibfnamefont {G.}~\bibnamefont {Wannous}}, \bibinfo {author}
  {\bibfnamefont {J.}~\bibnamefont {Vergès}},\ and\ \bibinfo {author}
  {\bibfnamefont {R.}~\bibnamefont {Barrow}},\ }\bibfield  {title} {\bibinfo
  {title} {Studies of the electronic states of the {BaF} molecule},\ }\href
  {https://doi.org/10.1080/00268979000101311} {\bibfield  {journal} {\bibinfo
  {journal} {Molecular Physics}\ }\textbf {\bibinfo {volume} {70}},\ \bibinfo
  {pages} {735} (\bibinfo {year} {1990})}\BibitemShut {NoStop}%
\bibitem [{\citenamefont {Bernard}\ \emph {et~al.}(1990)\citenamefont
  {Bernard}, \citenamefont {Effantin}, \citenamefont {D'Incan}, \citenamefont
  {Verg{\`{e}}s},\ and\ \citenamefont {Barrow}}]{Bernard1990}%
  \BibitemOpen
  \bibfield  {author} {\bibinfo {author} {\bibfnamefont {A.}~\bibnamefont
  {Bernard}}, \bibinfo {author} {\bibfnamefont {C.}~\bibnamefont {Effantin}},
  \bibinfo {author} {\bibfnamefont {J.}~\bibnamefont {D'Incan}}, \bibinfo
  {author} {\bibfnamefont {J.}~\bibnamefont {Verg{\`{e}}s}},\ and\ \bibinfo
  {author} {\bibfnamefont {R.~F.}\ \bibnamefont {Barrow}},\ }\bibfield  {title}
  {\bibinfo {title} {{Studies of the electronic states of the {BaF}
  molecule}},\ }\href {https://doi.org/10.1080/00268979000101321} {\bibfield
  {journal} {\bibinfo  {journal} {Molecular Physics}\ }\textbf {\bibinfo
  {volume} {70}},\ \bibinfo {pages} {747} (\bibinfo {year} {1990})}\BibitemShut
  {NoStop}%
\bibitem [{\citenamefont {Steimle}\ \emph {et~al.}(2011)\citenamefont
  {Steimle}, \citenamefont {Frey}, \citenamefont {Le}, \citenamefont {DeMille},
  \citenamefont {Rahmlow},\ and\ \citenamefont {Linton}}]{Steimle2011}%
  \BibitemOpen
  \bibfield  {author} {\bibinfo {author} {\bibfnamefont {T.~C.}\ \bibnamefont
  {Steimle}}, \bibinfo {author} {\bibfnamefont {S.}~\bibnamefont {Frey}},
  \bibinfo {author} {\bibfnamefont {A.}~\bibnamefont {Le}}, \bibinfo {author}
  {\bibfnamefont {D.}~\bibnamefont {DeMille}}, \bibinfo {author} {\bibfnamefont
  {D.~A.}\ \bibnamefont {Rahmlow}},\ and\ \bibinfo {author} {\bibfnamefont
  {C.}~\bibnamefont {Linton}},\ }\bibfield  {title} {\bibinfo {title}
  {Molecular-beam optical {Stark} and {Zeeman} study of the {$A$
  ${}^{2}\ensuremath{\Pi}$--$X$ ${}^{2}{\ensuremath{\Sigma}}^{+}$} (0,0) band
  system of {BaF}},\ }\href {https://doi.org/10.1103/PhysRevA.84.012508}
  {\bibfield  {journal} {\bibinfo  {journal} {Phys. Rev. A}\ }\textbf {\bibinfo
  {volume} {84}},\ \bibinfo {pages} {012508} (\bibinfo {year}
  {2011})}\BibitemShut {NoStop}%
\bibitem [{\citenamefont {Zhang}\ \emph {et~al.}(2022)\citenamefont {Zhang},
  \citenamefont {Zeng}, \citenamefont {Liang}, \citenamefont {Bu},\ and\
  \citenamefont {Yan}}]{Zhang2022}%
  \BibitemOpen
  \bibfield  {author} {\bibinfo {author} {\bibfnamefont {Y.}~\bibnamefont
  {Zhang}}, \bibinfo {author} {\bibfnamefont {Z.}~\bibnamefont {Zeng}},
  \bibinfo {author} {\bibfnamefont {Q.}~\bibnamefont {Liang}}, \bibinfo
  {author} {\bibfnamefont {W.}~\bibnamefont {Bu}},\ and\ \bibinfo {author}
  {\bibfnamefont {B.}~\bibnamefont {Yan}},\ }\bibfield  {title} {\bibinfo
  {title} {Doppler cooling of buffer-gas-cooled barium monofluoride
  molecules},\ }\href {https://doi.org/10.1103/PhysRevA.105.033307} {\bibfield
  {journal} {\bibinfo  {journal} {Phys. Rev. A}\ }\textbf {\bibinfo {volume}
  {105}},\ \bibinfo {pages} {033307} (\bibinfo {year} {2022})}\BibitemShut
  {NoStop}%
\bibitem [{\citenamefont {Hutzler}\ \emph {et~al.}(2012)\citenamefont
  {Hutzler}, \citenamefont {Lu},\ and\ \citenamefont {Doyle}}]{Hutzler2012}%
  \BibitemOpen
  \bibfield  {author} {\bibinfo {author} {\bibfnamefont {N.~R.}\ \bibnamefont
  {Hutzler}}, \bibinfo {author} {\bibfnamefont {H.-I.}\ \bibnamefont {Lu}},\
  and\ \bibinfo {author} {\bibfnamefont {J.~M.}\ \bibnamefont {Doyle}},\
  }\bibfield  {title} {\bibinfo {title} {{The Buffer Gas Beam: An Intense,
  Cold, and Slow Source for Atoms and Molecules}},\ }\href
  {https://doi.org/10.1021/cr200362u} {\bibfield  {journal} {\bibinfo
  {journal} {Chemical Reviews}\ }\textbf {\bibinfo {volume} {112}},\ \bibinfo
  {pages} {4803} (\bibinfo {year} {2012})}\BibitemShut {NoStop}%
\bibitem [{\citenamefont {Barry}\ \emph {et~al.}(2011)\citenamefont {Barry},
  \citenamefont {Shuman},\ and\ \citenamefont {DeMille}}]{Barry2011}%
  \BibitemOpen
  \bibfield  {author} {\bibinfo {author} {\bibfnamefont {J.~F.}\ \bibnamefont
  {Barry}}, \bibinfo {author} {\bibfnamefont {E.~S.}\ \bibnamefont {Shuman}},\
  and\ \bibinfo {author} {\bibfnamefont {D.}~\bibnamefont {DeMille}},\
  }\bibfield  {title} {\bibinfo {title} {A bright{,} slow cryogenic molecular
  beam source for free radicals},\ }\href {https://doi.org/10.1039/C1CP20335E}
  {\bibfield  {journal} {\bibinfo  {journal} {Phys. Chem. Chem. Phys.}\
  }\textbf {\bibinfo {volume} {13}},\ \bibinfo {pages} {18936} (\bibinfo {year}
  {2011})}\BibitemShut {NoStop}%
\bibitem [{\citenamefont {Bu}\ \emph {et~al.}(2017)\citenamefont {Bu},
  \citenamefont {Chen}, \citenamefont {Lv},\ and\ \citenamefont
  {Yan}}]{Bu2017}%
  \BibitemOpen
  \bibfield  {author} {\bibinfo {author} {\bibfnamefont {W.}~\bibnamefont
  {Bu}}, \bibinfo {author} {\bibfnamefont {T.}~\bibnamefont {Chen}}, \bibinfo
  {author} {\bibfnamefont {G.}~\bibnamefont {Lv}},\ and\ \bibinfo {author}
  {\bibfnamefont {B.}~\bibnamefont {Yan}},\ }\bibfield  {title} {\bibinfo
  {title} {Cold collision and high-resolution spectroscopy of buffer-gas-cooled
  baf molecules},\ }\href {https://doi.org/10.1103/PhysRevA.95.032701}
  {\bibfield  {journal} {\bibinfo  {journal} {Phys. Rev. A}\ }\textbf {\bibinfo
  {volume} {95}},\ \bibinfo {pages} {032701} (\bibinfo {year}
  {2017})}\BibitemShut {NoStop}%
\bibitem [{\citenamefont {Skoff}\ \emph {et~al.}(2009)\citenamefont {Skoff},
  \citenamefont {Hendricks}, \citenamefont {Sinclair}, \citenamefont {Tarbutt},
  \citenamefont {Hudson}, \citenamefont {Segal}, \citenamefont {Sauer},\ and\
  \citenamefont {Hinds}}]{Skoff2009}%
  \BibitemOpen
  \bibfield  {author} {\bibinfo {author} {\bibfnamefont {S.~M.}\ \bibnamefont
  {Skoff}}, \bibinfo {author} {\bibfnamefont {R.~J.}\ \bibnamefont
  {Hendricks}}, \bibinfo {author} {\bibfnamefont {C.~D.~J.}\ \bibnamefont
  {Sinclair}}, \bibinfo {author} {\bibfnamefont {M.~R.}\ \bibnamefont
  {Tarbutt}}, \bibinfo {author} {\bibfnamefont {J.~J.}\ \bibnamefont {Hudson}},
  \bibinfo {author} {\bibfnamefont {D.~M.}\ \bibnamefont {Segal}}, \bibinfo
  {author} {\bibfnamefont {B.~E.}\ \bibnamefont {Sauer}},\ and\ \bibinfo
  {author} {\bibfnamefont {E.~A.}\ \bibnamefont {Hinds}},\ }\bibfield  {title}
  {\bibinfo {title} {Doppler-free laser spectroscopy of buffer-gas-cooled
  molecular radicals},\ }\href {https://doi.org/10.1088/1367-2630/11/12/123026}
  {\bibfield  {journal} {\bibinfo  {journal} {New Journal of Physics}\ }\textbf
  {\bibinfo {volume} {11}},\ \bibinfo {pages} {123026} (\bibinfo {year}
  {2009})}\BibitemShut {NoStop}%
\bibitem [{Note1()}]{Note1}%
  \BibitemOpen
  \bibinfo {note} {See Ref.~\cite {Wright2022} for a study of this effect in
  various other molecular species.}\BibitemShut {Stop}%
\bibitem [{\citenamefont {Preston}\ \emph {et~al.}()\citenamefont {Preston},
  \citenamefont {Aufderheide}, \citenamefont {Ballard}, \citenamefont
  {Mawhorter},\ and\ \citenamefont {Grabow}}]{Mawhorter}%
  \BibitemOpen
  \bibfield  {author} {\bibinfo {author} {\bibfnamefont {A.}~\bibnamefont
  {Preston}}, \bibinfo {author} {\bibfnamefont {G.}~\bibnamefont
  {Aufderheide}}, \bibinfo {author} {\bibfnamefont {W.}~\bibnamefont
  {Ballard}}, \bibinfo {author} {\bibfnamefont {R.}~\bibnamefont {Mawhorter}},\
  and\ \bibinfo {author} {\bibfnamefont {J.-U.}\ \bibnamefont {Grabow}},\
  }\href@noop {} {\bibinfo {title} {(to be published)}}\BibitemShut {NoStop}%
\bibitem [{\citenamefont {Chen}\ \emph {et~al.}(2016)\citenamefont {Chen},
  \citenamefont {Bu},\ and\ \citenamefont {Yan}}]{Chen2016}%
  \BibitemOpen
  \bibfield  {author} {\bibinfo {author} {\bibfnamefont {T.}~\bibnamefont
  {Chen}}, \bibinfo {author} {\bibfnamefont {W.}~\bibnamefont {Bu}},\ and\
  \bibinfo {author} {\bibfnamefont {B.}~\bibnamefont {Yan}},\ }\bibfield
  {title} {\bibinfo {title} {Structure, branching ratios, and a laser-cooling
  scheme for the $^{138}\mathrm{BaF}$ molecule},\ }\href
  {https://doi.org/10.1103/PhysRevA.94.063415} {\bibfield  {journal} {\bibinfo
  {journal} {Phys. Rev. A}\ }\textbf {\bibinfo {volume} {94}},\ \bibinfo
  {pages} {063415} (\bibinfo {year} {2016})}\BibitemShut {NoStop}%
\bibitem [{\citenamefont {Brown}\ and\ \citenamefont
  {Carrington}(2003)}]{Brown2003}%
  \BibitemOpen
  \bibfield  {author} {\bibinfo {author} {\bibfnamefont {J.}~\bibnamefont
  {Brown}}\ and\ \bibinfo {author} {\bibfnamefont {A.}~\bibnamefont
  {Carrington}},\ }\href {https://books.google.de/books?id=BcZHngEACAAJ} {\emph
  {\bibinfo {title} {Rotational Spectroscopy of Diatomic Molecules}}},\
  Cambridge molecular science series\ (\bibinfo  {publisher} {Cambridge
  University Press},\ \bibinfo {year} {2003})\BibitemShut {NoStop}%
\bibitem [{\citenamefont {Drouin}\ \emph {et~al.}(2001)\citenamefont {Drouin},
  \citenamefont {Miller}, \citenamefont {Müller},\ and\ \citenamefont
  {Cohen}}]{Drouin2001}%
  \BibitemOpen
  \bibfield  {author} {\bibinfo {author} {\bibfnamefont {B.~J.}\ \bibnamefont
  {Drouin}}, \bibinfo {author} {\bibfnamefont {C.~E.}\ \bibnamefont {Miller}},
  \bibinfo {author} {\bibfnamefont {H.~S.}\ \bibnamefont {Müller}},\ and\
  \bibinfo {author} {\bibfnamefont {E.~A.}\ \bibnamefont {Cohen}},\ }\bibfield
  {title} {\bibinfo {title} {The rotational spectra, isotopically independent
  parameters, and interatomic potentials for the {$X\Pi_{3/2}$} and
  {$X\Pi_{1/2}$} states of {BrO}},\ }\href
  {https://doi.org/https://doi.org/10.1006/jmsp.2000.8252} {\bibfield
  {journal} {\bibinfo  {journal} {Journal of Molecular Spectroscopy}\ }\textbf
  {\bibinfo {volume} {205}},\ \bibinfo {pages} {128} (\bibinfo {year}
  {2001})}\BibitemShut {NoStop}%
\bibitem [{\citenamefont {Doppelbauer}\ \emph {et~al.}(2022)\citenamefont
  {Doppelbauer}, \citenamefont {Wright}, \citenamefont {Hofsäss},
  \citenamefont {Sartakov}, \citenamefont {Meijer},\ and\ \citenamefont
  {Truppe}}]{Doppelbauer2022}%
  \BibitemOpen
  \bibfield  {author} {\bibinfo {author} {\bibfnamefont {M.}~\bibnamefont
  {Doppelbauer}}, \bibinfo {author} {\bibfnamefont {S.~C.}\ \bibnamefont
  {Wright}}, \bibinfo {author} {\bibfnamefont {S.}~\bibnamefont {Hofsäss}},
  \bibinfo {author} {\bibfnamefont {B.~G.}\ \bibnamefont {Sartakov}}, \bibinfo
  {author} {\bibfnamefont {G.}~\bibnamefont {Meijer}},\ and\ \bibinfo {author}
  {\bibfnamefont {S.}~\bibnamefont {Truppe}},\ }\bibfield  {title} {\bibinfo
  {title} {Hyperfine-resolved optical spectroscopy of the {$A2\Pi$}
  $\leftarrow$ {$X2\Sigma+$} transition in {MgF}},\ }\href@noop {} {\bibfield
  {journal} {\bibinfo  {journal} {The Journal of Chemical Physics}\ }\textbf
  {\bibinfo {volume} {156}},\ \bibinfo {pages} {134301} (\bibinfo {year}
  {2022})}\BibitemShut {NoStop}%
\bibitem [{\citenamefont {Shaw}\ \emph {et~al.}(2021)\citenamefont {Shaw},
  \citenamefont {Schnaubelt},\ and\ \citenamefont {McCarron}}]{Shaw2021}%
  \BibitemOpen
  \bibfield  {author} {\bibinfo {author} {\bibfnamefont {J.~C.}\ \bibnamefont
  {Shaw}}, \bibinfo {author} {\bibfnamefont {J.~C.}\ \bibnamefont
  {Schnaubelt}},\ and\ \bibinfo {author} {\bibfnamefont {D.~J.}\ \bibnamefont
  {McCarron}},\ }\bibfield  {title} {\bibinfo {title} {Resonance {R}aman
  optical cycling for high-fidelity fluorescence detection of molecules},\
  }\href {https://doi.org/10.1103/PhysRevResearch.3.L042041} {\bibfield
  {journal} {\bibinfo  {journal} {Phys. Rev. Research}\ }\textbf {\bibinfo
  {volume} {3}},\ \bibinfo {pages} {L042041} (\bibinfo {year}
  {2021})}\BibitemShut {NoStop}%
\bibitem [{\citenamefont {Kajita}(2018)}]{Kajita2018}%
  \BibitemOpen
  \bibfield  {author} {\bibinfo {author} {\bibfnamefont {M.}~\bibnamefont
  {Kajita}},\ }\bibfield  {title} {\bibinfo {title} {Precise measurement of
  transition frequencies of optically trapped {40Ca19F} molecules},\ }\href
  {https://doi.org/10.7566/JPSJ.87.104301} {\bibfield  {journal} {\bibinfo
  {journal} {Journal of the Physical Society of Japan}\ }\textbf {\bibinfo
  {volume} {87}},\ \bibinfo {pages} {104301} (\bibinfo {year}
  {2018})}\BibitemShut {NoStop}%
\bibitem [{\citenamefont {Leung}\ \emph {et~al.}(2022)\citenamefont {Leung},
  \citenamefont {Iritani}, \citenamefont {Tiberi}, \citenamefont {Majewska},
  \citenamefont {Borkowski}, \citenamefont {Moszynski},\ and\ \citenamefont
  {Zelevinsky}}]{Leung2022}%
  \BibitemOpen
  \bibfield  {author} {\bibinfo {author} {\bibfnamefont {K.~H.}\ \bibnamefont
  {Leung}}, \bibinfo {author} {\bibfnamefont {B.}~\bibnamefont {Iritani}},
  \bibinfo {author} {\bibfnamefont {E.}~\bibnamefont {Tiberi}}, \bibinfo
  {author} {\bibfnamefont {I.}~\bibnamefont {Majewska}}, \bibinfo {author}
  {\bibfnamefont {M.}~\bibnamefont {Borkowski}}, \bibinfo {author}
  {\bibfnamefont {R.}~\bibnamefont {Moszynski}},\ and\ \bibinfo {author}
  {\bibfnamefont {T.}~\bibnamefont {Zelevinsky}},\ }\bibfield  {title}
  {\bibinfo {title} {A terahertz vibrational molecular clock with systematic
  uncertainty at the $10^{-14}$ level},\ }\href
  {https://arxiv.org/abs/2209.10864} {\bibfield  {journal} {\bibinfo  {journal}
  {arXiv:2209.10864}\ } (\bibinfo {year} {2022})}\BibitemShut {NoStop}%
\bibitem [{\citenamefont {Barontini}\ \emph {et~al.}(2021)\citenamefont
  {Barontini}, \citenamefont {Boyer}, \citenamefont {Calmet}, \citenamefont
  {Fitch}, \citenamefont {Forgan}, \citenamefont {Godun}, \citenamefont
  {Goldwin}, \citenamefont {Guarrera}, \citenamefont {Hill}, \citenamefont
  {Jeong}, \citenamefont {Keller}, \citenamefont {Kuipers}, \citenamefont
  {Margolis}, \citenamefont {Newman}, \citenamefont {Prokhorov}, \citenamefont
  {Rodewald}, \citenamefont {Sauer}, \citenamefont {Schioppo}, \citenamefont
  {Sherrill}, \citenamefont {Tarbutt}, \citenamefont {Vecchio},\ and\
  \citenamefont {Worm}}]{Barontini2021}%
  \BibitemOpen
  \bibfield  {author} {\bibinfo {author} {\bibfnamefont {G.}~\bibnamefont
  {Barontini}}, \bibinfo {author} {\bibfnamefont {V.}~\bibnamefont {Boyer}},
  \bibinfo {author} {\bibfnamefont {X.}~\bibnamefont {Calmet}}, \bibinfo
  {author} {\bibfnamefont {N.~J.}\ \bibnamefont {Fitch}}, \bibinfo {author}
  {\bibfnamefont {E.~M.}\ \bibnamefont {Forgan}}, \bibinfo {author}
  {\bibfnamefont {R.~M.}\ \bibnamefont {Godun}}, \bibinfo {author}
  {\bibfnamefont {J.}~\bibnamefont {Goldwin}}, \bibinfo {author} {\bibfnamefont
  {V.}~\bibnamefont {Guarrera}}, \bibinfo {author} {\bibfnamefont {I.~R.}\
  \bibnamefont {Hill}}, \bibinfo {author} {\bibfnamefont {M.}~\bibnamefont
  {Jeong}}, \bibinfo {author} {\bibfnamefont {M.}~\bibnamefont {Keller}},
  \bibinfo {author} {\bibfnamefont {F.}~\bibnamefont {Kuipers}}, \bibinfo
  {author} {\bibfnamefont {H.~S.}\ \bibnamefont {Margolis}}, \bibinfo {author}
  {\bibfnamefont {P.}~\bibnamefont {Newman}}, \bibinfo {author} {\bibfnamefont
  {L.}~\bibnamefont {Prokhorov}}, \bibinfo {author} {\bibfnamefont
  {J.}~\bibnamefont {Rodewald}}, \bibinfo {author} {\bibfnamefont {B.~E.}\
  \bibnamefont {Sauer}}, \bibinfo {author} {\bibfnamefont {M.}~\bibnamefont
  {Schioppo}}, \bibinfo {author} {\bibfnamefont {N.}~\bibnamefont {Sherrill}},
  \bibinfo {author} {\bibfnamefont {M.~R.}\ \bibnamefont {Tarbutt}}, \bibinfo
  {author} {\bibfnamefont {A.}~\bibnamefont {Vecchio}},\ and\ \bibinfo {author}
  {\bibfnamefont {S.}~\bibnamefont {Worm}},\ }\bibfield  {title} {\bibinfo
  {title} {{QSNET, a network of clocks for measuring the stability of
  fundamental constants}},\ }in\ \href {https://doi.org/10.1117/12.2600493}
  {\emph {\bibinfo {booktitle} {Quantum Technology: Driving Commercialisation
  of an Enabling Science II}}},\ Vol.\ \bibinfo {volume} {11881},\ \bibinfo
  {editor} {edited by\ \bibinfo {editor} {\bibfnamefont {M.~J.}\ \bibnamefont
  {Padgett}}, \bibinfo {editor} {\bibfnamefont {K.}~\bibnamefont {Bongs}},
  \bibinfo {editor} {\bibfnamefont {A.}~\bibnamefont {Fedrizzi}},\ and\
  \bibinfo {editor} {\bibfnamefont {A.}~\bibnamefont {Politi}}},\ \bibinfo
  {organization} {International Society for Optics and Photonics}\ (\bibinfo
  {publisher} {SPIE},\ \bibinfo {year} {2021})\ pp.\ \bibinfo {pages} {63 --
  66}\BibitemShut {NoStop}%
\bibitem [{\citenamefont {Athanasakis-Kaklamanakis}\ \emph
  {et~al.}(2023)\citenamefont {Athanasakis-Kaklamanakis}, \citenamefont
  {Wilkins}, \citenamefont {Breier},\ and\ \citenamefont
  {Neyens}}]{Athanasakis2023}%
  \BibitemOpen
  \bibfield  {author} {\bibinfo {author} {\bibfnamefont {M.}~\bibnamefont
  {Athanasakis-Kaklamanakis}}, \bibinfo {author} {\bibfnamefont {S.~G.}\
  \bibnamefont {Wilkins}}, \bibinfo {author} {\bibfnamefont {A.~A.}\
  \bibnamefont {Breier}},\ and\ \bibinfo {author} {\bibfnamefont
  {G.}~\bibnamefont {Neyens}},\ }\bibfield  {title} {\bibinfo {title}
  {King-plot analysis of isotope shifts in simple diatomic molecules},\ }\href
  {https://doi.org/10.1103/PhysRevX.13.011015} {\bibfield  {journal} {\bibinfo
  {journal} {Phys. Rev. X}\ }\textbf {\bibinfo {volume} {13}},\ \bibinfo
  {pages} {011015} (\bibinfo {year} {2023})}\BibitemShut {NoStop}%
\bibitem [{\citenamefont {Kogel}\ and\ \citenamefont
  {Langen}(2023)}]{Kogel2021code}%
  \BibitemOpen
  \bibfield  {author} {\bibinfo {author} {\bibfnamefont {F.}~\bibnamefont
  {Kogel}}\ and\ \bibinfo {author} {\bibfnamefont {T.}~\bibnamefont {Langen}},\
  }\href@noop {} {\bibfield  {journal} {\bibinfo  {journal} {in preparation}\ }
  (\bibinfo {year} {2023})}\BibitemShut {NoStop}%
\bibitem [{\citenamefont {Wright}\ \emph {et~al.}(2022)\citenamefont {Wright},
  \citenamefont {Doppelbauer}, \citenamefont {Hofsäss}, \citenamefont
  {Schewe}, \citenamefont {Sartakov}, \citenamefont {Meijer},\ and\
  \citenamefont {Truppe}}]{Wright2022}%
  \BibitemOpen
  \bibfield  {author} {\bibinfo {author} {\bibfnamefont {S.~C.}\ \bibnamefont
  {Wright}}, \bibinfo {author} {\bibfnamefont {M.}~\bibnamefont {Doppelbauer}},
  \bibinfo {author} {\bibfnamefont {S.}~\bibnamefont {Hofsäss}}, \bibinfo
  {author} {\bibfnamefont {H.~C.}\ \bibnamefont {Schewe}}, \bibinfo {author}
  {\bibfnamefont {B.}~\bibnamefont {Sartakov}}, \bibinfo {author}
  {\bibfnamefont {G.}~\bibnamefont {Meijer}},\ and\ \bibinfo {author}
  {\bibfnamefont {S.}~\bibnamefont {Truppe}},\ }\bibfield  {title} {\bibinfo
  {title} {Cryogenic buffer gas beams of {AlF}, {CaF}, {MgF}, {YbF}, {Al},
  {Ca}, {Yb} and {NO} – a comparison},\ }\href
  {https://doi.org/10.1080/00268976.2022.2146541} {\bibfield  {journal}
  {\bibinfo  {journal} {Molecular Physics}\ ,\ \bibinfo {pages} {e2146541}}
  (\bibinfo {year} {2022})}\BibitemShut {NoStop}%
\end{thebibliography}%

\end{document}